# MODELING AND ANALYSIS OF

# THE WIND-WAVES FIELD VARIABILITY

# IN THE INDIAN OCEAN DURING 1998-2009 YEARS

Polnikov V.G.[1)], Pogarskii F.A.[1)], Sannasiraj S.A.[2)], Sundar V.[2)]
[1)] Obukhov Institute for Atmospferic Physiscs of RAS, Moscow, Russia 119017
E-mail: polnikov@mail.ru
[2)] Indian Institute of Technology Madras, Chennai, India

## ABSTRACT

To calculate the wind-waves in the Indian Ocean (IO), the wind field for the period from 1998 to 2009 was used, obtained from the NCEP/NOAA archive [1], and numerical model WAM (Cycle-4) was applied, modified by the new source function proposed in [2]. Based on buoy data for the Indian Ocean, the fitting of the modified model WAM was done, which provides the win in accuracy of calculations on 35%, in comparison with the original model. All the further calculations of the wave fields in IO were made for these model settings.

At the first stage, the analysis of the simulation results involves a) mapping the fields of the significant wave height $<H_S(\mathbf{x},T,R)>$ and the wave energy $<E_W(\mathbf{x},T,R)>$, calculated with different scales of averaging in time $T$ and space $R$; b) estimating the fields of seasonal, annual and long-term variability; and c) determining the 12-year trend of the annually averaged fields, as well. The analysis was carried out taking into account the previously introduced zoning the ocean area, provided by the spatial inhomogeneity of the wind field [3].

Further analysis includes a) creation of time series for the averaged (over zones and across the ocean) wave height, $<H_S(R,t)>$, and wave energy, $<E_W(R,t)>$; b) construction of the frequency spectra of these series; c) finding the extrema of wave field; d) making histograms of wave heights (in the zones and the whole ocean); and e) calculating the first four statistical moments for the wave-height field (in the zones and whole IO).

The results obtained allow us to estimate the stored energy of the wave field in the Indian Ocean and the scales of its variability; to establish a positive 12-year trend of the averaged wave height (about 1% per year) and wave energy (2% per year); to determine features of the probability distribution; and to describe the statistical properties of the wave field in the zones of the Indian ocean.

Keywords: Indian Ocean, zone partition, wind-wave model, the wave field, the wave energy field, variability of the wave field, the spectra of the series, histogram of wave heights and their statistical moments.

## 1. Introduction and statement of objectives

This work is a continuation of implementation the international project of RFBR № 10-05-92 662-IND_a, devoted to study of an energy state of the wind and waves fields; processes of their interaction; and long-term variability on the example the Indian Ocean (IO). In a previous paper [3] it was carried out a comprehensive analysis of variability of the wind field in the IO area for the period 1998-2008yy; and the certain systematic approach to complex statistical processing of global geophysical fields was suggested. In this paper, this approach is used as a methodological standard for statistical analysis and studying variability of the wind-wave field in IO.

For constructing the field of wind waves, a wind field and a wind wave numerical model are required. The wind field used in this study was obtained from the site of NCEP/NOAA [1]. This field is a reanalysis performed on a grid with resolution of $1^0 \times 1.25^0$ on latitude-longitude and 3h in time. As it was established earlier [4], the accuracy of the field components is of the order of 1.5-2 m/s. At the time of the work execution, the available to us duration of the reanalysis was 12 years: from 1998 to 2009yy. Such a detailed and long-term wind field, to our knowledge, has never been used for detailed studying the wave field in the Indian Ocean (see references to alternative approaches in [3]).

In addition, a significant difference of the present work from similar studies is using the well-known numerical wind-wave model WAM (Cycle-4), in which the physical part of the model (the so-called source function) is replaced by the one proposed earlier in Polnikov [2]. As it was shown in special studies on the example of the Atlantic ocean [4-6], the referred replacement of the source function (both in WAVEWATCH [4] and in WAM [5,6]) leads to a substantial (15-20%) reducing the simulation errors for the wave field. For more convincing, in section 2 we present similar results obtained by comparing our simulation results with the data for three buoys in IO. It is these results (found in terms of using the modified model WAM) provide a great novelty and validity of the estimates of the wave field and its variability in IO, compared with the corresponding results of the global wave fields, which have been considered earlier [7, 8, 9 ].

The goals of this work do not relate to any detailed analysis of the previous research results of wave field studies in any regions of the World Ocean, as our project differs substantially from the well-known studies. In fact, it focuses mainly on the studying peculiarities of the space-time structure for the wind and wave energy fields, and, in the future, on the study of variability of their mechanical interaction on the long-term (climate) scale. Analogs of such studies are not known. Therefore, the above-mentioned papers are only the referring points, allowing to compare our results with those that overlap with the results of our works. This comparison is conducted at the conclusion of the article.

Additionally, it is essential to note the following distinctive feature of our study that is a detailed account of the spatial inhomogeneity of the wind field in the Indian Ocean, which is accomplished by a zoning the entire ocean to a number of areas that differ in dynamics of the wind field. As the basis of this zoning, we use the stable distributions of the local extrema for the averaged wind fields (see pictures below and in detail [3]). The introduction of zoning allows us identifying characteristics of the wind and waves fields variability with greater details (and credence) in comparison with the known results of earlier works.

With using the modified model WAM, the basic set of tasks of this paper is as follows.

1. To build and analyze the four types of charts of averaged fields for significant wave height $<H_S(i,j,T)>$, given by

$$<H_S(i,j,T)> = \left( \sum_{t_n \in T} \Delta t_n H_S(i,j,t_n) \right) / \sum_{t_n \in T} \Delta t_n \qquad . \qquad (1.1)$$

The kind of charts are as follows:

a) Charts for a winter (January) and a summer (July) month, made with averaging over all years;

b) The annually averaged charts;

c) Chart of the significant wave height averaged over the entire period;

d) Chart of the whole-period trend for the wave-height field in whole IO.

Hereafter, $\Delta t_n = \Delta t = 10800$s is the time step, $T$ is period of averaging, $H_S(i,j,t_n)$ is the significant wave height at the spatial node $(i, j)$ and at time $t_n$, given by the integral of the current two-dimensional spectrum of waves, $S(\omega,\theta,i,j,t_n)$:

$$H_S(i,j,t_n) = 4(\iint S(\omega,\theta,i,j,t_n) d\omega d\theta)^{1/2} \qquad , \qquad (1.2)$$

where $\omega$ and $\theta$ is the frequency and direction of the spectral components of the wave field, respectively.

Purposes are the following: to get a mean wave-height field $<H_S(i,j,T)>$ for the whole IO; to determine the seasonal and interannual variability of the mean field $<H_S(i,j,T)>$; to estimate the 12-year trend of the wave height field $H_S(i,j,t_n)$ (spatial distribution of the trend); basing on an analysis of the spatial distribution of wave heights, to control an extent of compliance the IO zoning for the wave and wind fields.

2. To build and analyze the charts for a mean field of wave-energy density, defined as

$$<E_W(i.j,T)> = \left( \rho_w g \sum_{t_n \in T} \Delta t_n H_S^2(i,j,t_n)/16 \right) / \sum_{t_n \in T} \Delta t_n \qquad (1.3)$$

where $\rho_w$ = 1029 kg/m$^3$ is the sea water density, and g = 9.81m/s$^2$ is the acceleration due to gravity. Variations of the argument $T$ correspond to the four types of averaging mentioned in Problem 1.

Purposes: to determine the average field of wave-energy density (hereafter, the wave energy field, for simplicity) in the IO area, in units $J/m^2$; its seasonal and annual variability; estimating the 12-year trend of the wave energy field and its spatial distribution.

3. To construct the time-history series of the wave height averaged in space and time:

$$H_S(R,T) = \left( \sum_{t_n \in T} \Delta t_n \sum_{i,j \in R} H_S(i,j,t_n) \Delta S_{ij} \right) / \sum_{t_n \in T} \Delta t_n / \sum_{i,j \in R} \Delta S_{ij} \qquad (1.4)$$

and the same series of the wave energy $E_W(R,T)$ defined as

$$E_W(R,T) = \left( \rho_w g \sum_{t_n \in T} \Delta t_n \sum_{i,j \in R} \frac{H_S^2}{16}(i,j,t_n) \Delta S_{ij} \right) / \sum_{t_n \in T} \Delta t_n / \sum_{i,j \in R} \Delta S_{ij} \qquad (1.5)$$

where $\Delta S_{ij}$ is the area of the special grid cell in the IO area, the left bottom corner of which is adjacent to the *i*-th node of latitude and the *j*-th node of longitude. Argument $R$ is the index of the spatial averaging region (i.e., a certain point, a zone, or the whole ocean), argument $T$ is the averaging period.

The following 12-year series are constructed: a) the time history for the "instantaneous" (3- hours time step) values of $H_S(Zn.,t)$ and $E_W(Zn,t)$ at the central points of the zones (denoted as $Zn$); b) the series $H_S(Zn,T,t)$ and $E_W(Zn,T,t)$ with the daily averaging ($T = 1$day) for each zone; and c) series of the daily-average values $H_S(R,T,t)$ and $E_W(R,T,t)$, obtained with additional spatial averaging both over the zones and over the entire IO.

The resulting time series are the objects for a spectral analysis.

Purposes: to determine the scales of temporal variability both the "instantaneous" wave height series $H_S(R,T,t)$ and the energy series $E_W(R,T,t)$, using data at the selected points, or data averaged over the zones, and over the whole IO.

4. To plot the time history of the annually averaged wave height $H_S(R,T,t)$ and wave energy $E_W(R,T,t)$ (for the whole period form 1998 to 2009) for each zone and for the whole IO.

Purpose: to determine mean values and the 12-year trends for annually averaged wave heights and energy both for zones and for whole IO.

5. To determine the extreme values of wave heights $H_{S,\max}(i_m, j_m, t_m)$ and their spatial and temporal location coordinates $(i_m, j_m, t_m)$ for each zone; to build a map of the extreme values of wave heights $H_{S,\max}(i,j)$ for the whole period.

The aim: to provide information of the modeled maximum wave heights values (the most likely close to real values), including their space and time distribution. This is important, for example, for the risk assessment of shipping.

6. To construct a histogram for the wave heights of the following types: a) at the central points of each zone, for the whole period from 1998 to 2009yy; b) for a chosen winter (January) and summer (July) month with accumulation for all the years (for each zone); c) the total histogram for the whole period, with the spatial accumulation for each zone separately and for whole IO.

The aim: to get the actual probability distribution function (PDF) of wave heights; to demonstrate their spatial and temporal variability within IO.

7. For all kinds of histograms, mentioned in item 6, to calculate the first four statistical moments (mean, standard deviation, skewness and kurtosis) and estimate the parameters of the modeling PDF, parameterized by the Weibull distribution.

The aim: to demonstrate the extent to which these PDF and statistical characteristics are close to those known from the literature [7-11].

According to the technique proposed in [3], the specificity of the present analysis is to study the statistics of the wave fields and their energy for the three types of spatial scales: at each point of the ocean (chart), the distribution of characteristics in the zones (series), and the integral characteristics of the entire ocean.[1] This approach can be roughly described as the principle of enlarging the scale describing a geophysical field. Farther, we will adhere to this principle. However, in view of the limited text volume, the graphic format will be given here for the most important results, only.

### 2. Assessing the accuracy of the modified model WAM

In this section we briefly review the results of a comparative verification of the modified model WAM, made on the basis of available to us data for three buoys located in the IO(Fig. 1). In this case, we will follow this process technology, developed and successfully tested previously [4-6].

Remind that the numerical model WAM is described by the transport equation [12]

$$\frac{dS(\sigma, \theta, \mathbf{x}, t)}{dt} = F(S, \mathbf{W}, \mathbf{U}) = In + Nl - Dis \qquad , \qquad (2.1)$$

in which the left hand side is the total derivative of the frequency-angular spectrum in time and the right hand side is the so-called source function F depending both on the spectrum of waves and on the local wind $\mathbf{W}(\mathbf{x}, t)$ and current $\mathbf{U}(\mathbf{x}, t)$.

---

[1] As we know, such approach to the statistical features description for geophysical fields is used for the first time(at least, for the IO case).

The left hand side of (2.1) is a mathematical part of the model. It is implemented in the WAM as a set of programs calculating the evolution of the wave spectrum in the spherical coordinates with using the standard initial and boundary conditions. In our calculations this part of the model remains unchanged.

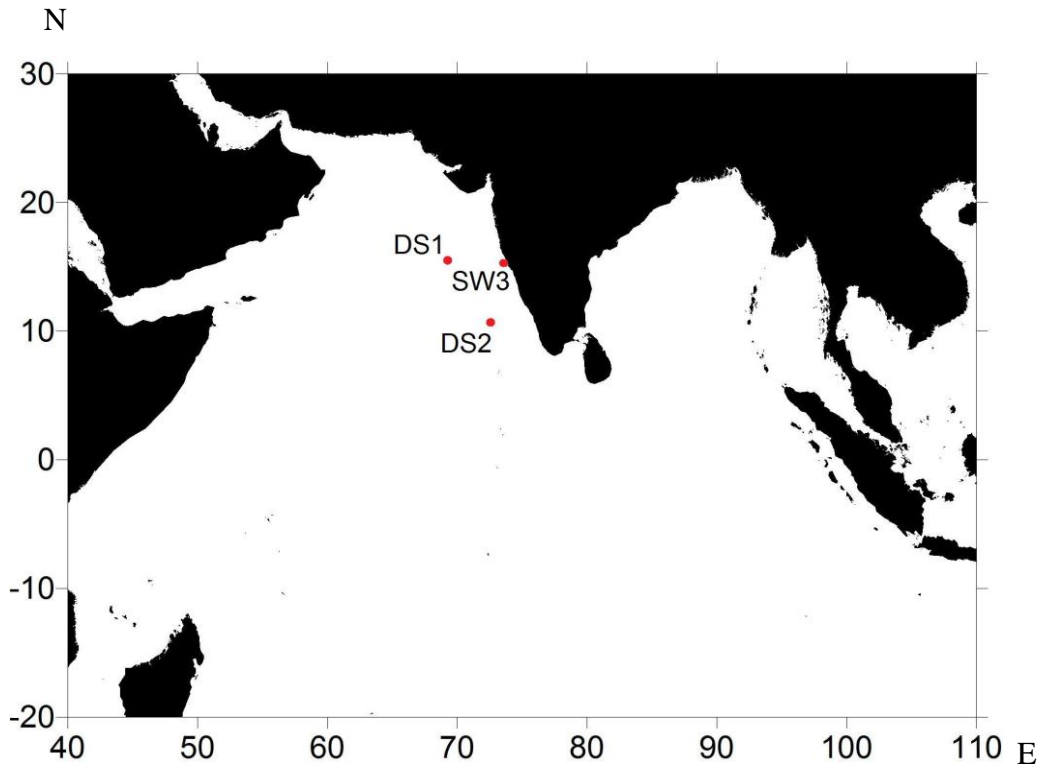

Fig. 1. The chart of buoy locations.

Mathematical representation for the terms of the source function (SF) deals with the physical content of the model. SF describes the mechanisms of wind-wave field evolution: a) the rate of energy transfer from wind to waves, *In*; b) the rate of nonlinear transfer of wave energy through the spectrum, *Nl*; and c) the rate of wave energy dissipation, *Dis*. In our work, just these components of the original model have been replaced by the others proposed in [2].

Detailed view of the modified SF, as well as the technology of performing a comparative verification are described previously [4-6], and we will not dwell on this issue. Therefore, we will present father the final estimations of the root-mean-square (r.m.s) errors for modeling values of significant wave height, $\delta H_S$, found by comparing them with those obtained at three buoys located in the Arabian Sea (see Fig. 1 and Tab. 1) [2].

As seen from the Tab. 1, the win in accuracy of the modified model, i.e. the decrease of $\delta H_S$, amounts, on average, about 35% , with respect to the errors of the original model. In the terminology of paper [6], this is an indication of the significant improvement of the model. In our calculations, the mean value $\delta H_S$ for the modified model is about 0.35m.

---

[2] A more detailed description of the verification procedure for the modified WAM, based on the buoy data in IO, will be presented in a separate paper.

Table 1.

The coordinates of 3 buoys in IO and r.m.s. errors of simulations for wind wave $H_S$ with two versions of WAM

| Index of buoy [coordinates] | $\delta H_S$, m WAM-orig | $\delta H_S$, m WAM-modif |
|---|---|---|
| DS1 [15.5N, 69.3E] | 0.75 | 0.47 |
| DS2 [10.7N, 72.5E] | 0.40 | 0.29 |
| SW3 [15.4N, 73.7E] | 0.45 | 0.28 |
| Mean r.m.s. error | 0.533 | 0.347 |

The result of calculations comparing, on the example of buoy DS1, is shown visually in Fig. 2. It is seen that the main gain in accuracy for the modified model is provided by more adequate description of the extreme values of wave heights (explanation of this fact can be found in [4,6]). This fact suggests that the results of calculations of extreme wave heights in IO, obtained by using the modified model, have higher reliability compared to those obtained with the original models WAM (or WW[4]).

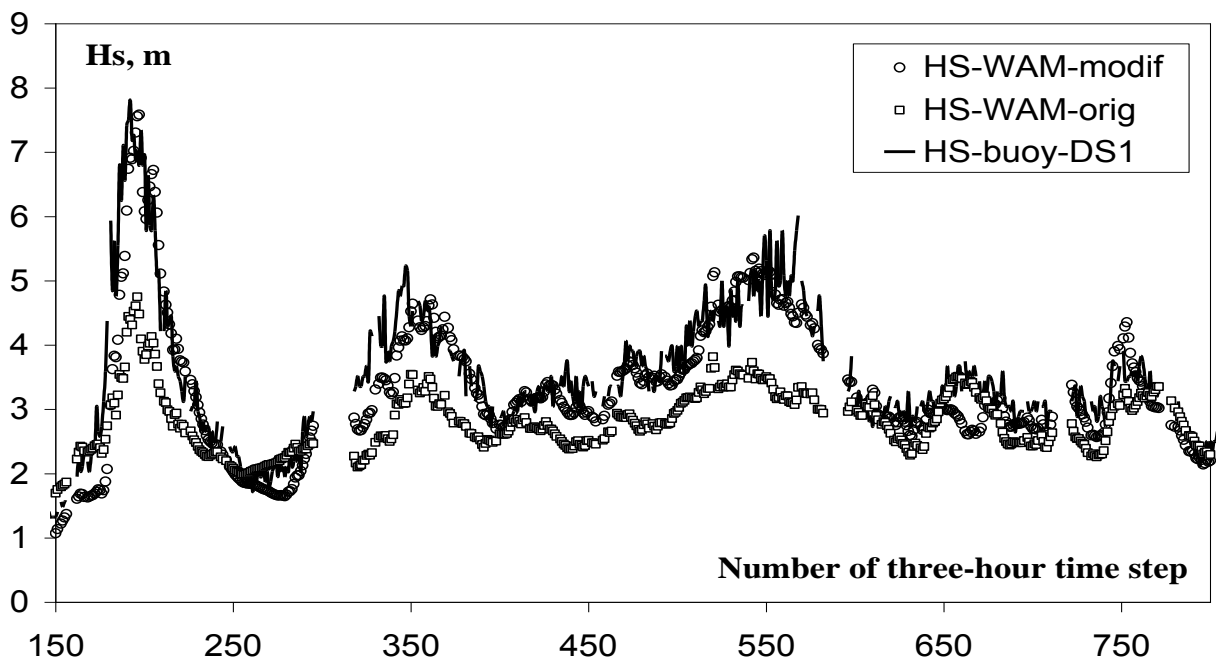

Fig. 2. Time history of the significant wave height $H_S(t)$ at the point of buoy DS1 location for the period of observations: the solid line is the buoy data; the line with open circles is the result of simulation with the original model WAM(C4); the line with filled circles is the result of simulation with the modified WAM.

## 3. Analysis of charts for the fields of waves and their energy

*3.1. Charts of the averaged fields of wave height*

From the variety of charts for the averaged fields of wave heights $<H_S(i,j,T)>$, said in the statement of the problems, the three most important charts are presented for demonstration (Fig. 3a, b, c). The mean January and mean July maps of $<H_S(i,j,T)>$ are given in Fig.-1pp; the mean 1998 and mean 2009 map are given in Fig. 2-app, in Appendix. The general analysis of the whole variety of maps shows the following features of wave-height fields $<H_S(i,j,T)>$.

Firstly, all the charts show a noticeable and sustainable spatial distribution of the wave heights, as shown on the example of chart for wave heights averaged for the whole 12-year period (Fig. 2a). As in the wind field case, we can see six separated zones (designated as Zn), characterized by the local extrema of wave height $H_S$: the Arabian Sea - Z1, the Bay of Bengal - Z2, the Equatorial part of IO - Z3, the trade-wind part of the south IO - Z4, the southern part of the subtropical IO - Z5, and the South Indian Ocean - Z6. The reason of this partition is provided by the dynamics of the wind field, which was discussed in details in [3]. That permits us do not dwell on this issue. It is important to note that the fact of sustainable spatial inhomogeneity of the wave height field takes place for all the secondary fields: seasonal, annual, and averaged for the whole period.

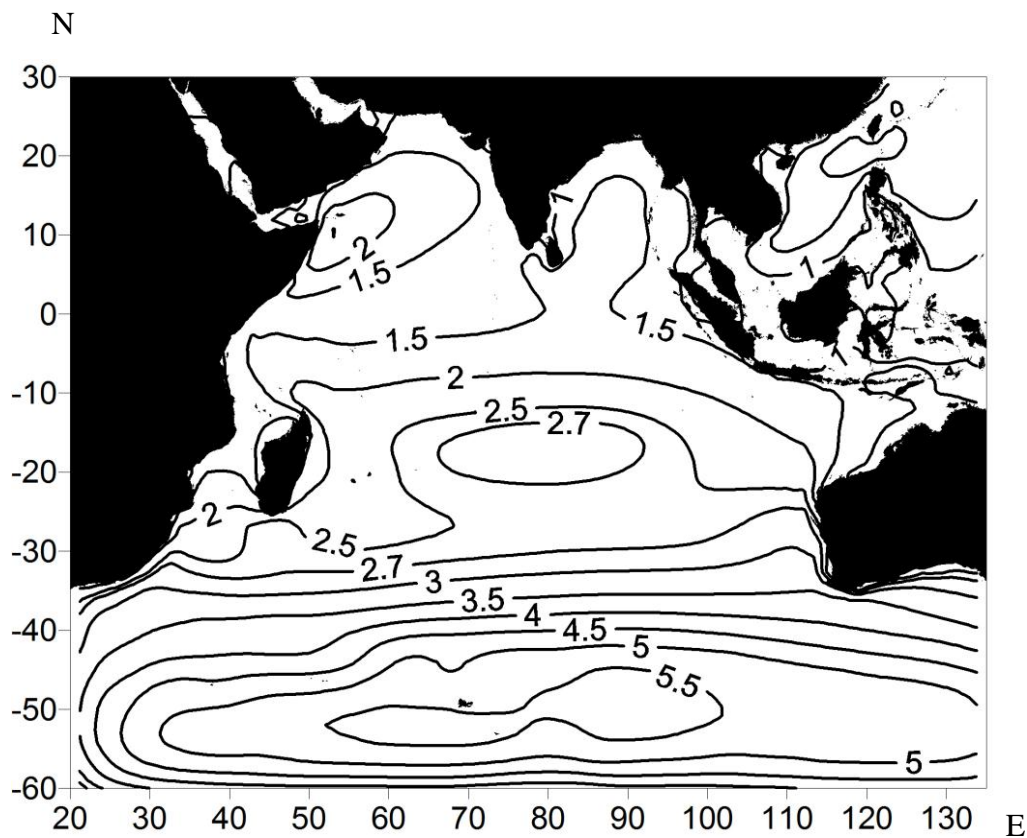

3a

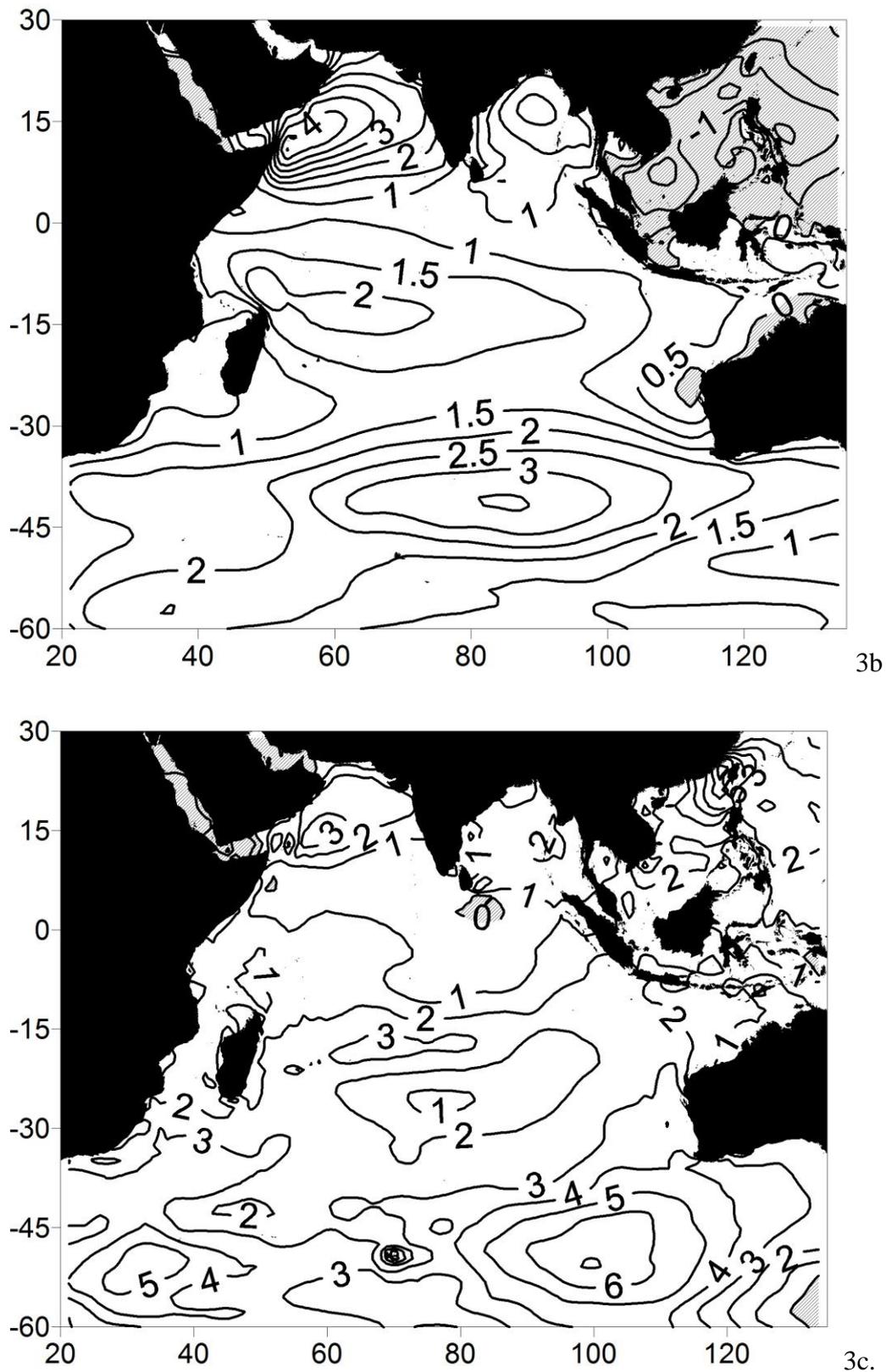

Fig.3. Charts of simulated fields: a) the field of wave height $H_S$ averaged for the whole period of 1998-2009yy, contours are given in meters; b) the field of seasonal difference (July-January) for the monthly fields averaged for the whole period, given in meters; c) the field of r.m.s. trend for wave fields $H_S$, given in sm/year. Shading corresponds to regions with negative trends.

Note that because of inertia of the wave field in comparison with the variability of the wind, the zone boundaries for the wave field are somewhat shifted with respect to those for wind fields, and in the wave fields they are not always clearly identified (see Table 2). However, in order of continuity the analysis established in [3], and taking into account the insignificance in the zones boundaries changes, farther we will use the previously adopted zoning partition of IO, which is given in Tab. 2 for completeness[3].

The second feature of the wave-wind field concerns determining the characteristic values of the fields. In the order of the primary analysis of charts shown in Fig. 3, note that the average for 12 years values $H_S$ are varying from the minimum values of 1m, in the coastal areas of zones, to the maximum value of the order of 5.5-6m, in the center of zone Z6 (Fig. 3a). Herewith, the maximum wave is often realized in the summer (Fig. 3b). Note that unlike the 12-year-averaged fields, the seasonal- and annual-averaged fields, saving the same zone structure, show a greater range of variability. Discussion of them and the exact values of wave heights, averaged over all ocean zones, will be given later (Section 5).

Table 2.

Partition of the wind field in IO and its manifestation in the wave field

| Zone | Coordinates of the zone boundaries for the wind field | Central points of the wind field zones | Extent of the zone manifestation in the wave field |
|---|---|---|---|
| Z1 | 40E - 80E; 25N - 7N | 14N, 62,5E | middling |
| Z2 | 80E - 100E; 25N - 7N | 15N, 88,75E | clearly (smartly) |
| Z3 | 35E - 105E; 7N - 9S | 00N, 80,0E | middling |
| Z4 | 35E - 142E; 9S – 22S | 17S, 80,0E | clearly (smartly) |
| Z5 | 20E - 140E; 22S –35S | 29S, 80,0E | poorly |
| Z6 | 20E - 147E; 35N -60N | 49S, 80,0E | clearly (smartly) |

Third, it is important to note a very strong seasonal variability of the mean field $<H_S(i,j,T)>$, shown on the example of the difference "summer-winter" fields averaged for 12 years (Fig. 3b). Almost everywhere, except the eastern part of the zone Z1, the summer field has an increase in the average wave heights. Especially this increase is noticeable in the western part of zone Z2 and at the center of the border between zones Z5 and Z6. This is different from the location of maxima for the seasonal variability of wind field (Fig. 1b in [3]). Summer growth of wave heights may reach 3-4m, i.e., about 100% of the average winter values. In this case, the negative seasonal changes of wave heights in the IO in the summer are practically not observed, except a narrow region near the north-western Australian coast. Thus, the monsoon variability of the wind, which is a source of the seasonal variations of the wave field, is the most pronounced only in the north-western part of the IO.

---

[3] Remind, that the most southern boundary is provided by the commonly used boundary between Indian and Southern oceans[3], which is taken as a conventional mean ice-boundary in the case of wind-wave simulation.

And finally, some words about the trend of wave heights, averaged for the whole period. As seen from a comparison of Figs. 3a and 3b, the regularity of the averaged heights distribution in zones is practically repeated in the field of its time-trend, obtained by the method of least squares for each point of the wave field with a resolution 3h ("instantaneous field"). Compared with the same trend of the wind field, the trend-field for wave heights is characterized by a greater smoothness, what is determined by the wave field inertia.

For the whole 12-year period, the average wave height has an increase with the rate of 1% per year (which exceeds the limits of the confidence interval for wave-height variability, $\Delta_H$, having a value of about 0.3-0.5%). The trend is most pronounced in areas of high winds (zone Z2, Z4, Z6). It is interesting to note that the magnitude of the mean trend of "instantaneous" wave fields of 1% corresponds to the trend of the values $H_S$ averaged across the ocean (see Section 5). At the same time, the chart of Fig. 2c has also a small region of negative trend, which is located on the western border of zones Z1 and Z3. However, this negative trend does not exceed 0.1% per year, well below the confidence intervals of these estimates. Interesting that, in contrast to the wind field, a negative trend for the east coast of Australia is not detected. Obviously, for a more reliable determination of the negative trends of wave heights (and their explanations), it needs a longer period of data analysis and a special discussion.

### 3.2. *Charts of wave energy*

Charts of wave energy, $<E_w(i.j,T)>$, are the obvious derivatives of the wave-height fields $<H_S(i,j,t_n)>$. This means that the averaged fields of wave energy repeat the wave height fields $<H_S(i,j,T)>$, emphasizing greatly the areas of extreme waves. However, calculation and analysis of variability in the wave-energy fields has its own justification.

The importance of wave energy fields, as well as the fields of wind energy, is determined by a large physical content of the energy concept compared with the concepts of wave height and wind speed. For this reason, as the basis of our research, we put to study a spatial distribution of wave and wind energy and their variability, taking into account that these fields determine the intensity of mechanical energy transfer from wind to waves, the detailed analysis which we plan to perform in subsequent papers. The said indicates a high degree of self-sufficiency and worth of analysis of the wave energy field. To our knowledge, there are no analogues of such calculations and their analysis in the literature.

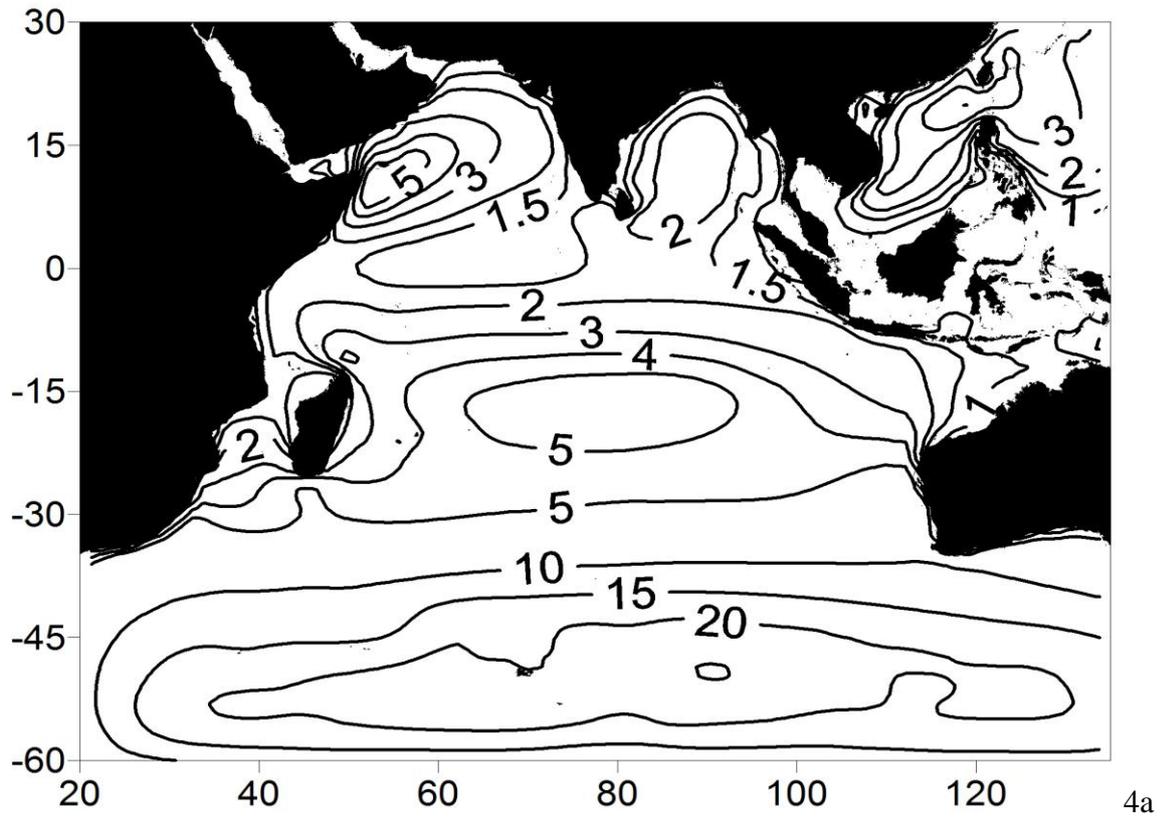

4a

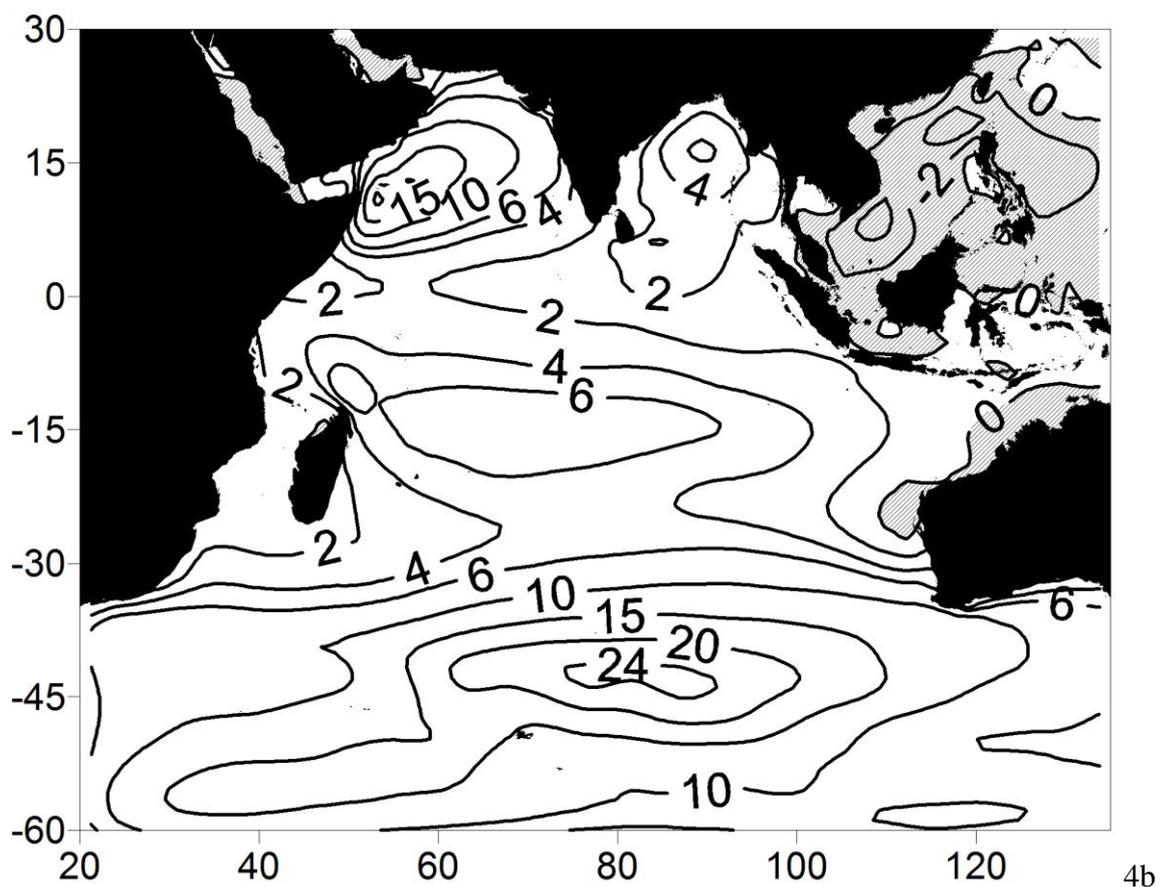

4b

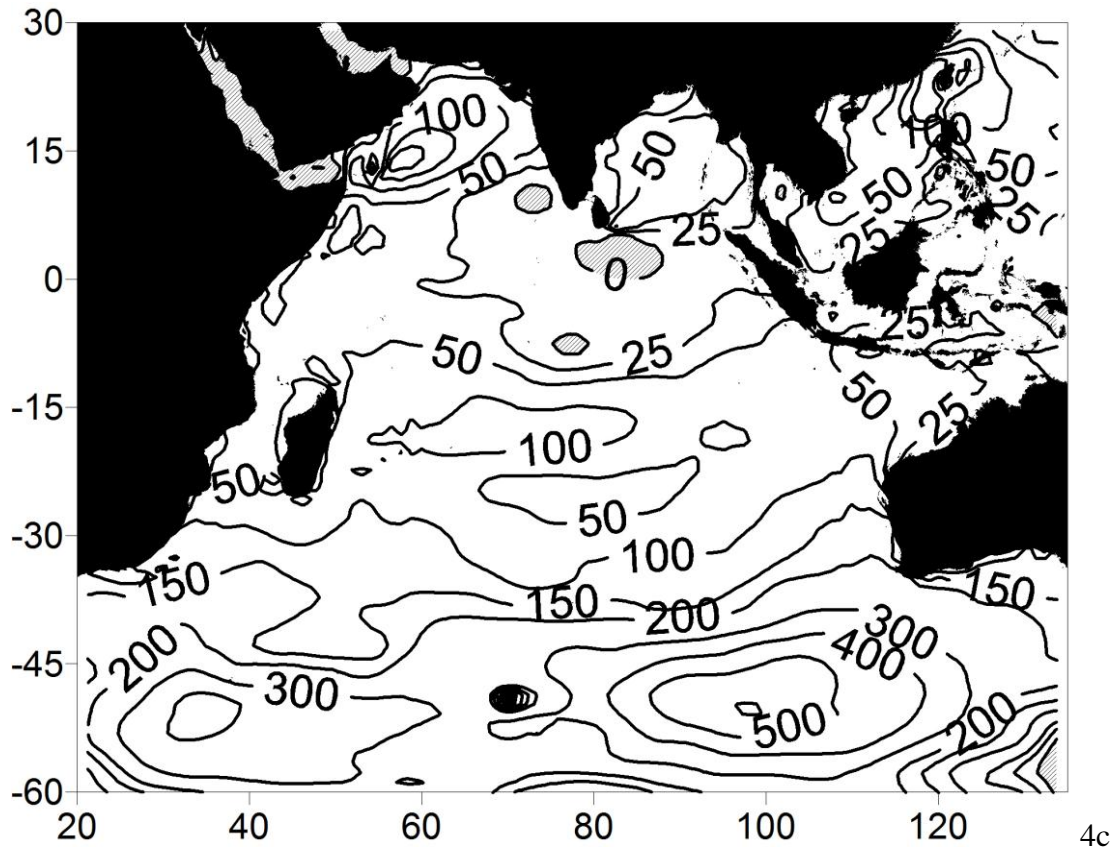

Fig. 4. The same as in Fig. 3 but for the wave energy.
Panels a) and b) are given in KJ/m$^2$ panel c) is given in J/m$^2$.

The most important results of calculations of the energy fields of the waves are shown in Figs 4a, b, c. In comparison with the fields of wind energy, the results indicate the following features of the wave energy fields $<E_w(i.j,T)>$.

First of all, note that the previously established zone-structure of the fields is preserved under any scale of averaging time (seasons, years, entire period), and their spatial distribution become more exaggerated. So, for the field $<E_w(i.j,T)>$, averaged for the 12-year period, the range of variability gets a factor 20, i.e. 4-5 times higher that of a similar field $<H_S(i,j,T)>$ (see Figs. 3a and 4a). At the peak, the characteristic value of wave energy averaged for 12 years reaches a value of 20 KJ/m$^2$, while the minimum values are of the order of units of KJ/m$^2$ (in the coastal areas of IO and in the open ocean of zone Z3). These values agree well with those for the fields $<H_S(i,j,T)>$, noted above.

The above consistency for geometry of fields $<H_S(i,j,T)>$ and $<E_w(i.j,T)>$ is seen for all scales of time averaging. Therefore, in view of the foregoing detail description of the wave height fields $<H_S(i,j,T)>$, the further description of the energy fields $<E_w(i.j,T)>$ requires some elaboration, only.

In particular, to describe the temporal variability of the wave energy field it is sufficient to give quantitative estimates of its seasonal trends and long-term variability.

The range of seasonal, "summer-winter" variability (Fig. 4b) changes from 25KJ/m$^2$ (at the maximum, corresponding to the center of zone Z6) to 2J/m$^2$ (at the minimum, which covers the entire equatorial zone Z3). These estimates are well correlated with the quantities of the seasonal variability of wave heights (Fig. 3b), and indicate about of 100% increase of wave energy in the summer over whole IO, practically.

The interannual 12-year trend of "instantaneous" wave energy fields, produced by the method of least squares at each point of the wave energy field for the whole period of calculations, has a range of variation from zero (at the border of zones Z1-Z2-Z3 and to the south of the Hindustan peninsula) to values of 500J/m$^2$ in the center of zone Z6 (Fig. 4c). Thus, during this period the long-term trend is about 2% per year (on average for the entire period). Herewith, the main increase in energy occurs in the western and eastern parts of zone Z6. Zero trend in the region located to the south of Hindustan is an unexpected feature of the field $<E_w(i.j,T)>$, justification of which is to be found by analysis of the overall variability of the atmospheric circulation dynamics. It is not excluded that this is a specificity of the long-term variability of the wave field, the study of which is still far from its completion and requires further versatile efforts.

Comparing the values of fluxes for wind kinetic-energy (about $10^3$W/m$^2$) with the values of the wave energy density (of the order $10^4$J/m$^2$) and using the estimate that the relative rate of energy flux from wind to waves is of the order of $(2-3)10^{-6}$ [13] from the current wind energy flux, one can obtain that "the full pumping" of the seen wave field in the whole IO by the wind (in the stationary case) can be realized for time about $10^6$ seconds, i.e. for 1-2 days. This result indicates a fairly rapid energy transfer dynamics of wind-waves interaction (in view of the ocean scale). It will be shown (Section 5) that such dynamics is actually realized, although the period of full pumping is few larger (7-10 days).

Summarizing, we note that the above estimates are absolutely new results in studying the wave field variability of IO. It is natural to expect that the work in this topic requires its continuation, which may result in some refinement of these estimates.

To the end of the wave energy field study, it should be noted that for purposes of comparison, the wind and waves energy-fields variability, the following issues are of the interest: a) the spatial distribution of local extrema for wind and wave energy fields both for instantaneous and for seasonal (and, possibly, for annually averaged) fields; b) the local extrema values of the seasonal and interannual trends of these fields and the variability of their locations; and c) the time scales of variability in these fields. Therefore, a detailed description of these items of a joint studying the spatial and temporal variability of fields $<E_w(i.j,T)>$ и $<E_A(i.j,T)>$, obviously, requires a farther separate discussion.

4. **Analysis of the time variation of wave heights and energy**. **The scales of their variability**

*4.1. The time history of wave heights at fixed points of zones (3-hours time step)*

Consider, first of all, the time history of "instantaneous" wave heights $H_S(Z_I,t)$ (step in time is 3h) at the central points of zones of IO, $Z_I$, the coordinates of which are given in Tab. 2. This approach is important to determine the full scale of temporal variability of the wave field and its variations, while moving from the north to the south. The proper figures are given in Appendix, Fig. 3-app. First we give a verbal description of the series, which will be enforced by the spectral analysis of them, given below.

The visual analysis of the 3h-series shows a number of differences between the wave and wind fields. The main one is that in all areas the wave field has the annual harmonic as the most pronounced (while in the time series of wind field, in zones Z1-Z3 the semiannual harmonics are clearly visible, see [3]). At the same time, in the northern zones of the wave field (as in the fields of wind) it is clearly visible the variability on the scale of tens days, what is replaced by the smaller scale variability while moving to the south. These observations are fully confirmed by spectral processing of the series, done on the basis of the autoregressive Yule-Walker analysis [14] (Fig. 5). Note that according to the formulas of [14], the confidence intervals for the spectra on a logarithmic scale is of the order of (15-20)%, which corresponds to the amplitudes of rapid intensity oscillations of the spectrum curves, visually broadening the plot-line at high frequencies ($f > 0.1$, where the axis of frequency $f$ is given in units of inverse days).

From the analysis of the frequency spectra $S(f)$ shown in Fig. 5a, b, we can draw the following conclusions. For zones Z1-Z3 having a similar wind and wave dynamics (Fig. 5a), the period of main variability is equal to 1 year. Then, with increasing the frequency, the spectral curve is continued by the slowly decaying in intensity cascade of the weakly expressed scales from 30 to 10 days. Beginning with a period of 10 days, the spectrum takes the power falling form

$$S(f) \propto f^{-n} \qquad (4.1)$$

with the relatively high index $n \cong 3.5 - 4$. Such a high degree of the spectrum decay (compared with the values of *n* for wind, in the same frequency band) is evidently due to the nonlinear dependence of wave heights on wind (see, for example, the ratio of [6]). In addition, it is important to note the pronounced peaks of the spectrum of wave heights on scales of 1, 0.5 and 0.3 days, repeating the same extremes in the spectrum for wind [3].

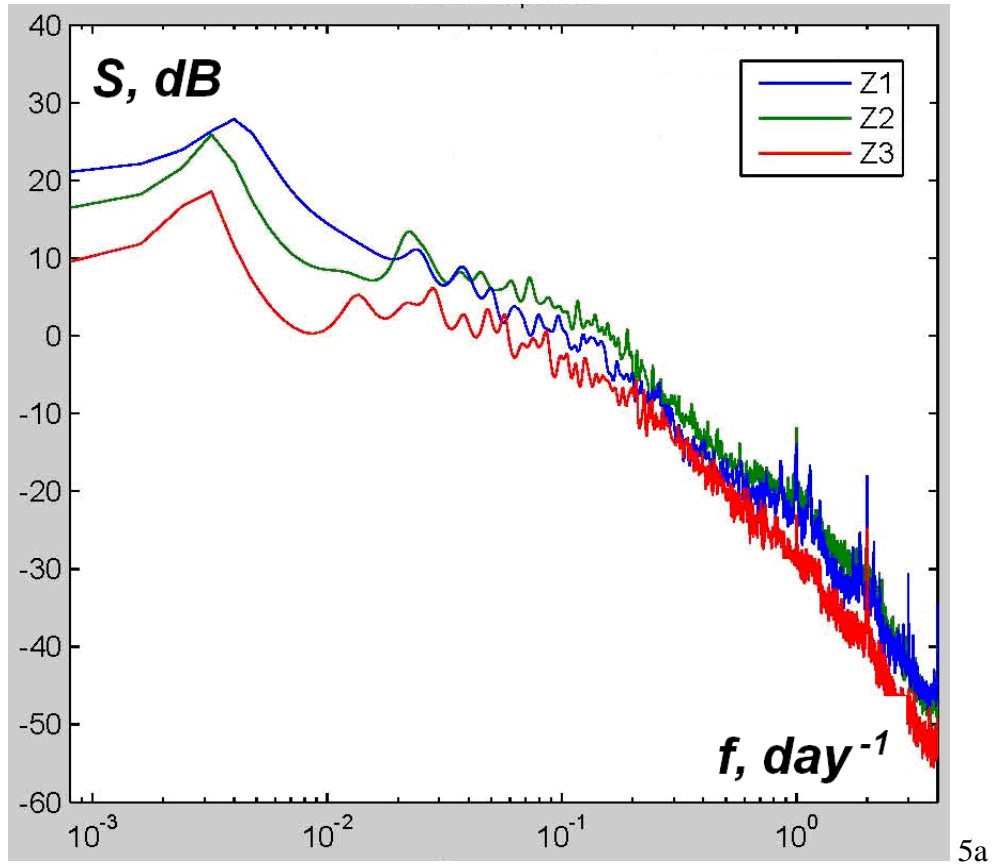

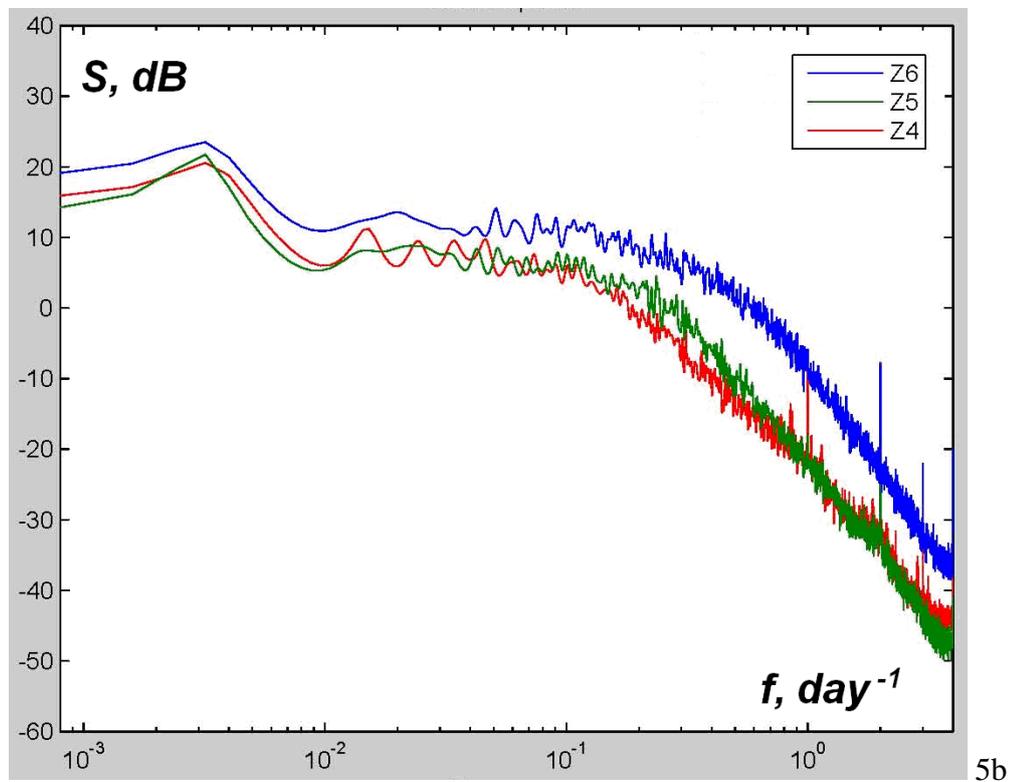

Fig. 5. Spectra of 3h-series for the wave height at the central point of zones in IO:
a) zones Z1- Z3; b) zones Z4 - Z6.

For the southern zone Z4-Z6, the wave height spectra are qualitatively similar to ones for the northern zones (Fig. 5b), but the cascade of scales get the form of a "shelf" of the white noise spectrum. This shelf occupies a scale from 100 to 10 days and repeats the same form for the spectrum of wind. As proposed earlier [3], such a shelf is a consequence of the lack of correlation of wind variability on these time scales, which naturally affects the variability of wave heights.

*4.2. Time history of the wave energy at fixed points of zones (3h-step)*

It is interesting to compare the above results with their counterparts for the series of wave energy $E_W(Z_I,t)$. In this case, since the series $E_W(Z_I,t)$ repeats the time dependence of wave heights, $H_S(Z_I,t)$, increased in amplitudes, the main interest is not the series but their spectra.

The calculations showed that the spectra of wave energy have the form similar to the spectra of heights, differing mainly by the magnitude of their intensity (therefore not shown). In this respect, the wave energy spectra are less interesting than the spectra of wind energy, which showed a larger set of variability scales (see [3]). Apparently, this "poor" set of scales for the wave energy spectra is due to the higher inertia of their field, smoothing some of the smaller scales of variability for the wind energy field.

Another remarkable difference between wave energy spectra and spectra for wave heights, in addition to their intensity, is the increasing degree of the spectra decay in the range of periods from 10 to 0.25 days (Nyquist frequency) on 15-20%. For the energy spectra $S(f)$, the decay exponents of the form (4.1) have values n $\cong 4.5-5$.

The said above gives a complete description of the wave energy series and their spectra, what allows do not dwell on more details, shifting this issue for a special consideration.

*4.3. The spectra of wave heights and energy, averaged for one day and over the zones*

Analysis of the data of this kind reveals a significant difference between the spectra of wave heights averaged for a day and over zone and the spectra of "instant" heights. This difference consists in the following (Fig. 6). As seen from Fig. 6a, for these series in the northern zones Z1-Z3, apart from the prevailing scale of variation of one year, a marked semiannual variability takes place, which is similar to one for wind fields [3]. While moving to the south, intensity of this variability is waning, almost disappearing in zones Z4-Z6 (Fig.6b).

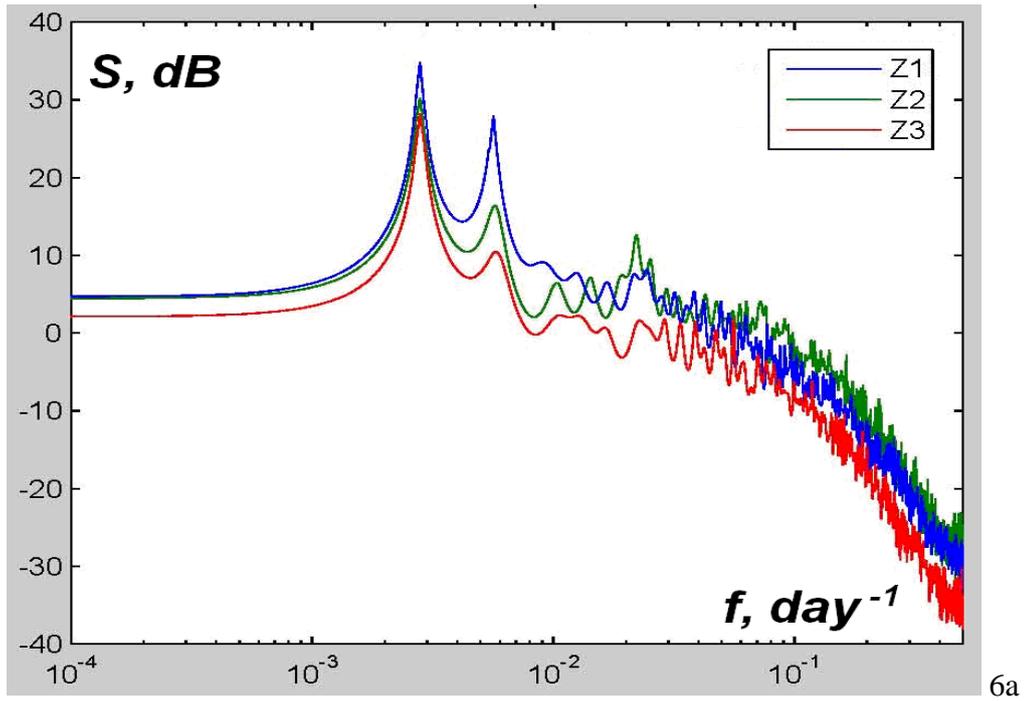

6a

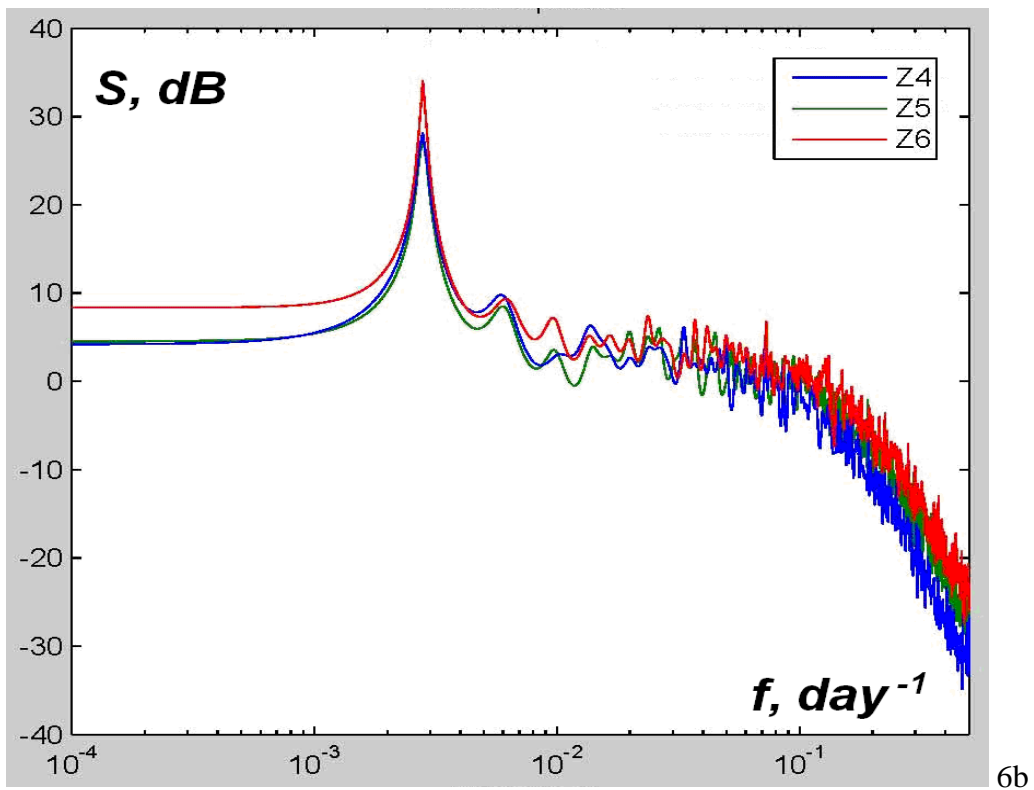

6b

Fig. 6. Spectra of the series for the daily-mean wave height averaged over the zones:
a) zones Z1- Z3; b) zones Z4 - Z6.

This behavior indicates the synchrony of semiannual variability of wave heights in the zones as a whole, which become apparent in the spectra of the averaged series. Though, at certain points of zones, as we have seen, this scale of variability is almost not seen (within the confidence intervals).

For the northern zones, there are clearly visible a credible intensity peaks on the 30-day scale. In the southern zones, similar peaks already occupy a large range of scales, appearing on the background of a long "shelf" located in scales from 100 to 10 days. These scales of variability are not very pronounced, however, they deserve to mention them. It is not excluded that they are the multiple harmonics of the annual fluctuations, provided by the nonlinear nature of wind waves. However, in general, the transfer of energy on the scale from 100 to 10 days is not seen, what is characterized by the "shelf" of the spectrum intensity in this frequency range. Clarification of the nature of these features for the spectrum of the averaged series of wave heights does obviously need a further study.

Regarding to the slopes of the spectra in the high-frequency band (for the periods less than 10 days), these slopes do not differ from those for the "instantaneous" wave heights. Such behavior of spectra for averaged series demonstrates the similarity of the dynamics of waves (alike of wind) on these scales[4].

In the final of this subsection, we note that the spectra of series for wave energy, averaged for a day and over zones, practically do not differ from those for the series of averaged heights. Therefore, they are not discussed here, for brevity.

*4.4. The time history and spectra of wave heights and energy, averaged for a day and over IO*

Consider now the features of the time history and the spectra of the daily-mean values $H_S(R,T,t)$ and $E_W(R,T,t)$, averaged over the entire IO (Figs. 7 and 8).

First, regarding the series $H_S(R,T,t)$ and $E_W(R,T,t)$, it needs to say that, despite the large scale of spatial averaging, these series show a very strong variability. Thus, on a scale of several days (5-10days), the average wave height of the ocean can vary up to 1.5-2 times, and the average energy does to a factor 3-4 (Fig. 7a). Such a strong variability of wave energy for the whole ocean is seemed as a much unexpected result, as far as it testifies to the intensive wave dynamics of IO. Moreover, such variability is typical for the whole period under consideration (Figure 7b). Herewith, it is even visually seen a tendency of growth of the wave energy averaged over an annual period (Fig. 7b).

Some estimates, explaining the possibility of "complete wave-pumping" IO by the wind on scales of 1-2 days, have been given previously in section (3.2). Here, this effect is confirmed visually in the series $H_S(R,T,t)$ и $E_W(R,T,t)$ representing values averaged over the ocean. However, the model data gives 3-4 times greater duration for the "complete ocean pumping", what is a significant refinement of the purely theoretical estimates made in section (3.2). It is clear that the increase of the wave pumping duration is due to the strong heterogeneity and non-stationarity of the wind field. The value and importance of the above series is that they give a more realistic assessment of the proper scale.

---

[4] Naturally that the 1-day and smaller periods are not expressed in the spectrum, as far as they are averaged in such series.

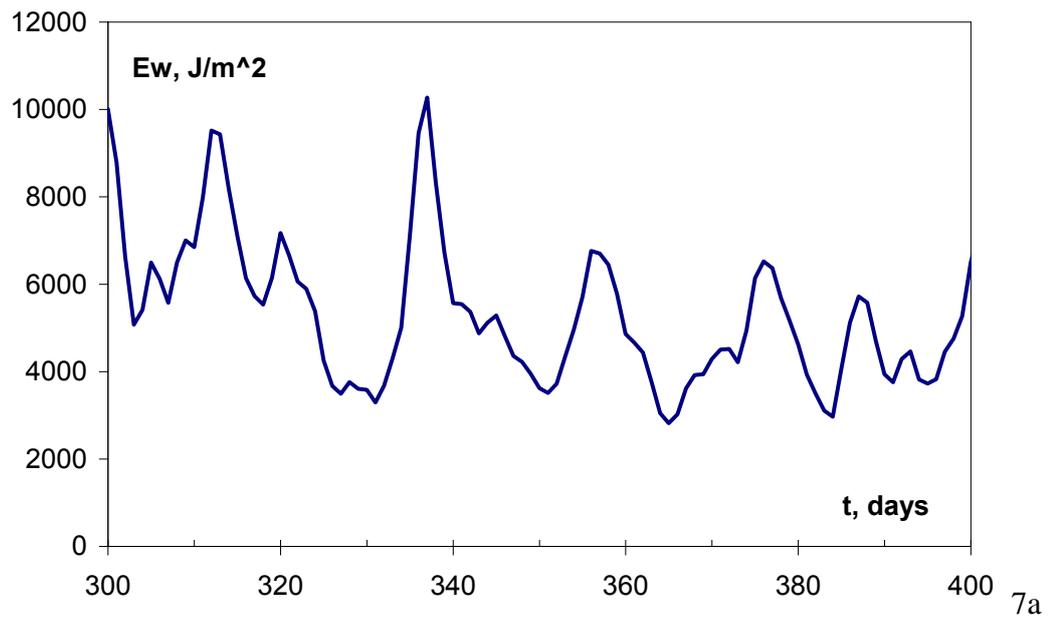

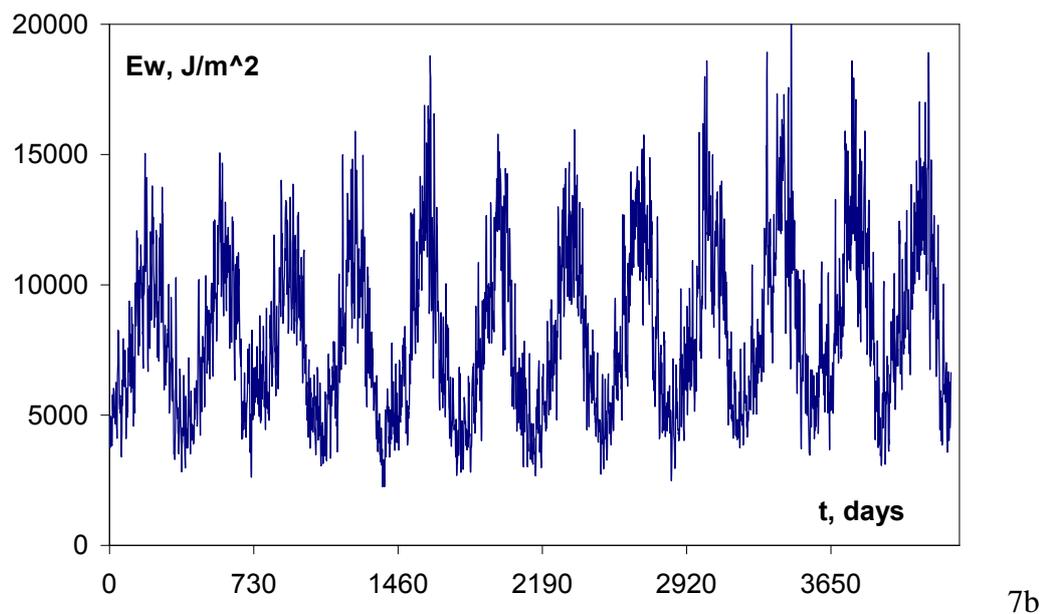

Fig. 6. The series of wave energy averaged for a day and over the whole IO:
a) detailed representation of the series; b) general kind of the series.

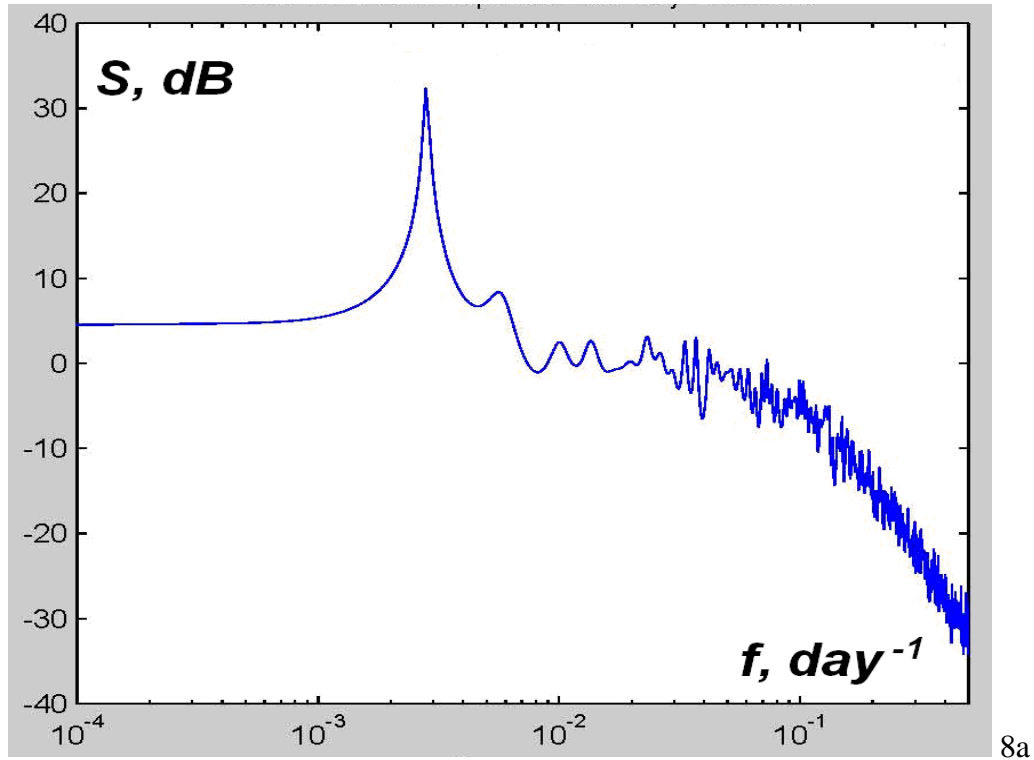

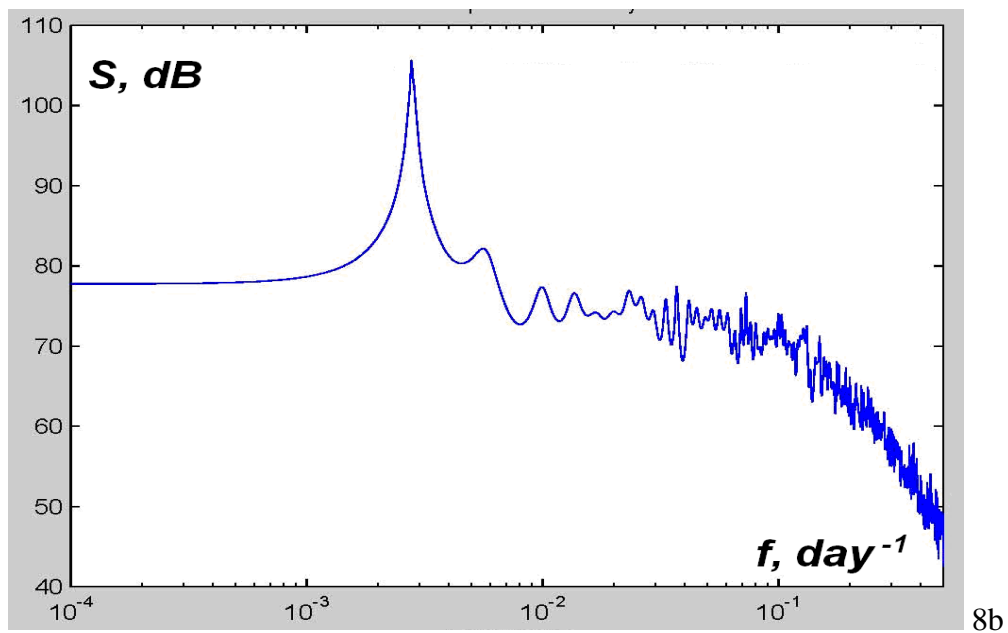

Fig. 7. Spectra of the series for the daily-mean values averaged over whole IO:
a) wave height; b) wave energy.

Second, regarding to the spectra. The spectrum of the wave heights averaged over IO (Fig. 8a) has a single well-expressed maximum corresponding to the 1-year period, on the background of which the semiannual harmonic is practically negligible, although its intensity exceeds the confidence intervals. In this respect, the wave-heights spectra differ from of the spectra of wind energy, in which the semi-annual harmonic was significant even for the whole IO[3].

Further, with increasing frequency, there is the already known "shelf" of intensity of the spectrum, extended from periods of 100 days to 10 days, on the background of which the spectrum peaks corresponding to scale of 20, 15 and 10 days take place (these scales are visually seen in Fig. 7a, as well). Starting from periods of 10 days, the slowly decaying spectrum is changed by the rather sharp power-law decay of the form (4.1), having exponent $n \cong 3$. At the same time, on the tail of the spectrum the selected scales of 8-7 days and 5-3 days are slightly visible, though they are well manifested in the spectrum of wind [3].

As to the wave energy spectrum for series of data averaged for a day and across the ocean (Fig.8b), this spectrum is fully similar in shape to the spectrum of averaged wave heights shown in Fig. 8a. It is presented here with the aim to show its difference form the analogous spectrum for wind energy, which has a wider set of variability scales [3].

### *4.5. Conclusions of the section*

The above analysis of the series and spectra for the wave heights and wave energy, in our opinion, has no analogues. In view of presentation diversity of time series for the wave field, the results shown above are preliminary and not exhaustive. The issue of their detailed description requires a separate and more comprehensive study and discussion. Nevertheless, even here we can formulate a number of conclusions.

*4.5.1.* First, the spectra of wave heights and wave energy, found on any scales of averaging in time and space, have the similar shapes. They differ in the intensity and rate of decay for the high-frequency tails of spectrum, only. In this regard, the wave energy spectra are not such diverse in the expressed scales as the spectra of wind energy [3].

*4.5.2.* Both for wave heights and for their energy, the spectra of the "instantaneous" series with 3h-step have the following scales of variability: 1-year, the cascade of scales from 30 to 10 days, dropping down in intensity, and scales of 1 day, 0.5 days, and 0.3 days. Starting from the scales of 10 days (or less), the spectra has the power-law decay of the form (4.1) with the exponents *n* equal to 3.5-4 for wave heights, and to 4-5 for their energy.

*4.5.3.* In the southern zone Z4-Z6, the characteristic feature of the spectra is a "shelf" in intensity, which occupies the scales from 100 to 10 days. This feature repeats the same singularity of the wind spectra[3]. The reason for this peculiarity is the lack of correlation in the variability for the wind field, and hence for the wave field, at these scales in the southern part of the IO. In the northern zones, such correlation is seemed to be partially introduced by processes of the monsoon dynamics of the wind field. This hypothesis requires the further study.

*4.5.4.* For daily values averaged over the zones, in the spectra of the both variables, the clearly have two well-expressed peaks corresponding the variability scales of 1 year and 0.5 year. The set of scales of the order of 40-20 days is seen as well. The intensity of the six-month period decreases significantly while moving to the south, reaching the level of the confidence intervals in zone Z6. At the same time, the variability intensity on the scales of the order of 40-20 days shows a small change, remaining remarkable the background of "shelf", typical for the southern zones, that occupies the scale from 100 to 10 days. Apparently, the scales of 40-20 days correspond to the recurrence of the large cyclones passage through the IO.

*4.5.5.* For the daily values average over the ocean, the spectra shapes for wave heights and wave energy are practically similar to the spectra of the series averaged over zone Z6 (see Section 4.5.4), due to a large space-proportion of this zone. Therefore, the spectra for such series have a white noise kind, i.e. "shelf", located between the scales of maximum variability and the small scales of turbulent nature. Interpretation of this feature is given in sub-section 4.5.3.

*4.5.6.* On the scales smaller than the white noise scale, all kinds of spectra have the power-law decay of type (4.1). Apparently, this is due to that the dynamics of wind processes on smaller scales (turbulent transfer of energy through a cascade of scales) is similar in all zones. This leads to conservation the falling spectrum at these scales for any variants of the fields averaging, both for the wind field and for the wave field. Exponents of the decay law are given in sub-section *4.5.2*.

**5. Analysis of integral and local features**

*5.1. The time history of annual values of wave heights and wave energy*

On two panels of Fig. 9 the time history of annual values averaged over zones and IO are shown for wave heights (Fig. 9a) and wave energy (Fig. 9b). It can be seen the quite expected distribution of average values among the zones. It is interesting to note that the estimates of annually averaged energy in the zones (with an error of about 10%) correspond to the square of the average wave height multiplied by 1000 of the kind

$$E_W(R,T) \cong 1000 H_S^2(R,T). \tag{5.1}$$

This empirical fact can be used for quick "expert" estimates of the average energy of waves, without involving cumbersome calculations by formula (1.5).

Limiting by the analysis of integral quantities for whole IO, it should be noted the presence of a well distinguished trend for the average wave height: a growth about 1% her year, what is at the border of the confidence intervals of the series variability. This trend is accompanied by a more pronounced trend for the ocean-averaged wave energy: an increase of 2% per year, which already exceeds the confidence intervals, and can be considered as a reliable result. The same conclusion is following from estimation of $R^2$ statistical measure ($R^2$= 0.67 for the average wave height (Fig.9a), and $R^2$= 0.77 for wave energy (Fig. 9b)).

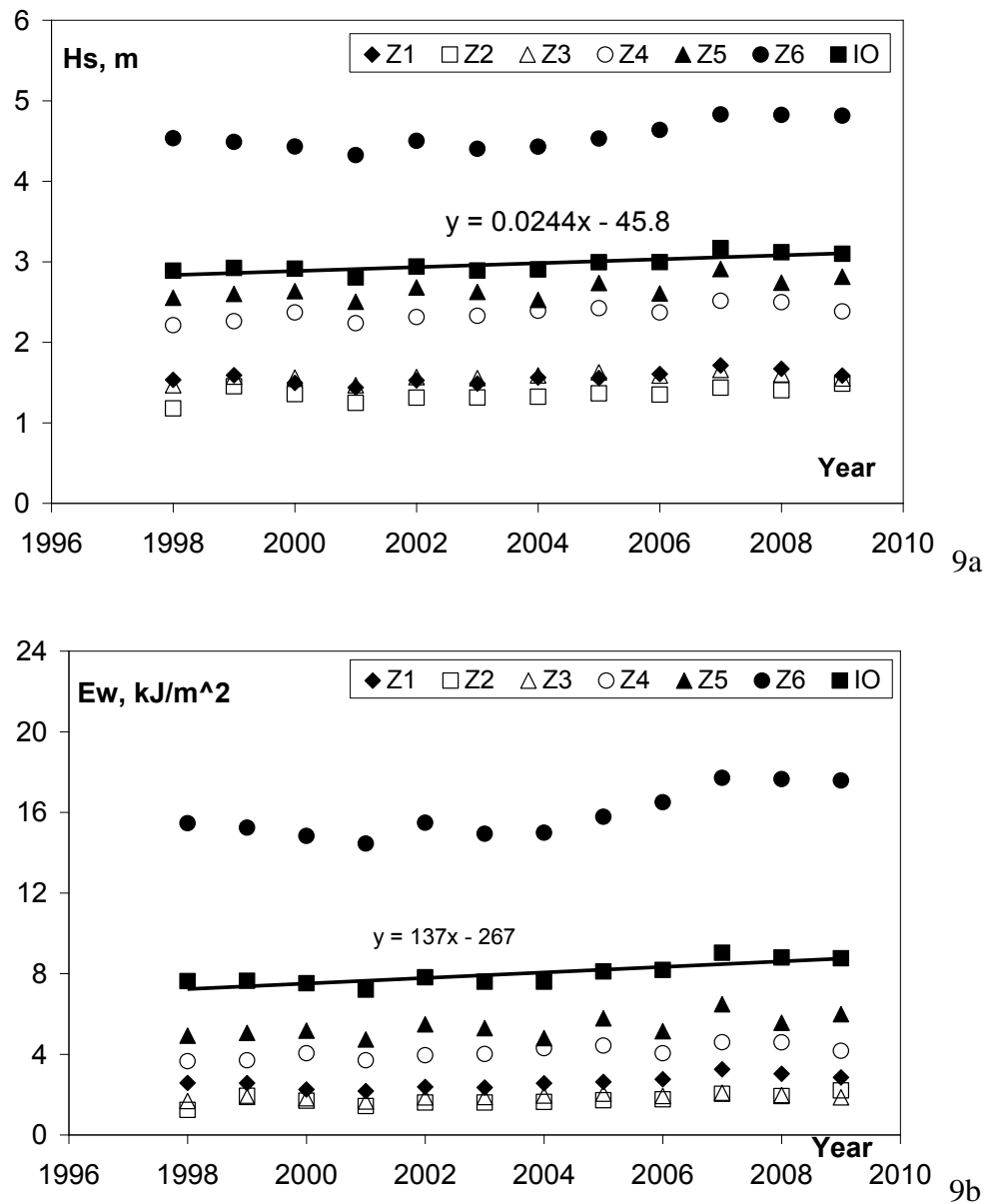

Fig. 9. Time history of the annually-mean values averaged over the zones and whole IO:
a) wave height ($R^2$= 0.67) ; b) wave energy ($R^2$ = 0.77 ).
Shown are the lines and formulas of r.m.s. trends for the case of averaging over whole IO.

The value of 2%-per year trend for the average ocean wave energy is of major interest in the light of the assessing (determination) the sought long-term variability of wave energy in the IO area. It is natural that in view of smallness of the time period (12 years), the obtained estimates have a local feature (i.e., not a climate feature). Nevertheless, they are important both for their clarification, valid on more long periods of consideration, and for purposes of comparison of the wind energy trends with the similar trends of wave energy, representing the main interest of the project. In this case, we have a quite reasonable correspondence between these quantities: the wind energy trend obtained earlier is about 1% per year [3] what correlates well with an increase in wave height of 1% per annum and wave energy growth of 2% per year.

It should be emphasized that the detailed numerical calculations of energy of the wave field in IO of such kind does not have analogues. Therefore, we believe that these estimates are of substantial interest both scientific and practical. In more details this issue is discussed in the concluding section.

### 5.2. Location of wave heights extremes

The values of the wave heights extrema, $H_{S\max}$, and coordinates of their spatial and temporal distribution, $(i_m, j_m, t_m)$, are shown in Tab. 3. Especially note that these space-time coordinates are of interest to compare them with the similar coordinates of the wind speed maxima $W_{\max}$ [3]. Such a comparison is of fundamental importance both for the demonstration of the difference itself and to assess the extent of scattering the extremes for the wind and wave fields.

A joint analysis of the wind and wave maxima tables testifies the following.

First, the largest magnitude of wave heights, reaching the values of 20-22m, is realized in the southern zone of IO: Z6, as might be expected from an analysis of the wind fields. However, spatial-temporal positions of the extrema $H_{S\max}$ are not quite the same as for $W_{\max}$ both in zone Z6 and in all other zones. This important fact, found by numerical methods, indicates that the instantaneous field of wave and wind differ significantly in topology. This means that the wave field cannot be assessed from an analysis of wind field, and the former needs an independent calculation. In particular, the calculations revealed that 2007 year has the greatest number of extrema, while the maximum number of extremes of wind in zones of IO was in 1998.

Secondly, the absolute maximum $H_{S\max}$, which is equal to 23.2 m, takes place in the "roaring" fortieth (coordinates: 49S, 86E), as well as its analogue $W_{\max}$. However, the time coordinates of these two events are separated significantly: in the case of wind field, the absolute extreme is realized on the date of 21.06.1998, while in case of wave field, it does on the date of 21.09.2009.

Third, in zones Z1 and Z3, 12-year maxima $H_{S\max}$ reach everywhere the value of the order of 10m, while in zone Z2, it is greater: 12m. In zones Z4 and Z5, wave extremes can reach values of 16-17m, while in the zone Z6, they are more than 22m. These digits show that, despite the apparent lack of tropical cyclones on the grid of wind fields with their very high values (as indicated by the analysis of wind fields, made in [3]), in the fields of wave heights, the extreme values $H_{S\max}$ are represented well enough. This fact is due to the inertia of wave field, the height of which is determined not only by the

Table 3.

Values and space-time locations of $H_{S\max}$ through zones

| \ZONE Year\ | 1 | 2 | 3 | 4 | 5 | 6 |
|---|---|---|---|---|---|---|
| 1998 | **9.0**<br>13.0\| 67.50<br>14/12 | **11.2**<br>21.0\| 87.50<br>10/07 | **8.8**<br>0.0\|47.50<br><u>21/05</u> | **14.7**<br>-19.0\| 58.75<br>10/02 | **13.6**<br>-35.0\| 76.25<br>14/02 | **18.2**<br>-54.0\| 72.50<br>03/12 |
| 1999 | **8.2**<br>11.0\| 56.25<br>07/07 | **8.1**<br>19.0\| 88.75<br>17/06 | **9.8**<br>-9.0\| 77.50<br>20/10 | **9.8**<br>-12.0\| 72.50<br><u>08/10</u> | **12.3**<br>-35.0\| 52.50<br>17/05 | **16.5**<br>-59.0\|116.25<br>20/08 |
| 2000 | **7.7**<br>12.0\| 52.50<br>03/07 | **8.8**<br>10.0\| 83.75<br>25/12 | <u>**10.6**</u><br><u>-9.0\| 82.50</u><br><u>05/01</u> | **11.4**<br>-21.0\|102.50<br>03/03 | **12.1**<br>-35.0\|131.25<br>21/08 | **17.4**<br>-58.0\| 76.25<br>12/10 |
| 2001 | **7.4**<br>12.0\| 53.75<br>14/07 | **7.9**<br>16.0\| 88.75<br>12/06 | **7.1**<br>-9.0\| 62.50<br>03/04 | **11.6**<br>-12.0\|112.50<br>12/02 | **14.6**<br>-35.0\| 82.50<br>28/07 | **18.1**<br>-54.0\| 47.50<br>29/07 |
| 2002 | **8.1**<br>12.0\| 57.50<br>14/06 | **9.0**<br>17.0\| 92.50<br><u>18/05</u> | **6.7**<br>-9.0\| 47.50<br>17/09 | **9.4**<br>-19.0\| 77.50<br>06/02 | **13.8**<br>-35.0\| 72.50<br>17/05 | **21.2**<br>-57.0\| 67.50<br>31/07 |
| 2003 | **8.2**<br>11.0\| 53.75<br>25/07 | **8.2**<br>16.0\| 92.50<br><u>23/07</u> | **6.6**<br>-9.0\| 56.25<br>10/02 | **10.7**<br>-22.0\| 66.25<br>13/03 | **13.5**<br>-35.0\| 57.50<br>31/07 | **19.3**<br>-47.0\| 37.50<br>29/08 |
| 2004 | **9.9**<br>16.0\| 66.25<br>14/06 | **8.3**<br>17.0\| 86.25<br>13/06 | **8.9**<br>-9.0\| 72.50<br>24/11 | **11.6**<br>-17.0\| 51.25<br>07/03 | **14.3**<br>-35.0\|108.75<br>24/08 | **18.3**<br>-43.0\|126.25<br>29/06 |
| 2005 | **8.7**<br>13.0\| 56.25<br>24/07 | **8.2**<br>16.0\| 87.50<br>24/07 | **6.7**<br>-8.0\| 87.50<br>12/10 | **9.3**<br>-22.0\| 57.50<br>24/03 | **15.3**<br>-35.0\| 87.50<br>13/05 | **19.5**<br>-55.0\| 91.25<br>28/09 |
| 2006 | **9.5**<br>15.0\| 58.75<br>01/07 | **9.1**<br>16.0\| 87.50<br>02/07 | **7.3**<br>-9.0\| 51.25<br>22/12 | **8.7**<br>-19.0\|113.75<br>21/01 | **13.2**<br>-35.0\| 62.50<br>03/05 | **20.1**<br>-54.0\|111.25<br>09/10 |
| 2007 | <u>**10.6**</u><br><u>12.0\| 57.50</u><br><u>17/06</u> | <u>**12.6**</u><br><u>16.0\| 87.50</u><br><u>28/06</u> | **8.4**<br>6.0\| 52.50<br>18/06 | <u>**16.5**</u><br><u>-22.0\| 52.50</u><br><u>27/02</u> | <u>**17.5**</u><br><u>-30.0\| 51.25</u><br><u>01/03</u> | **22.7**<br>-41.0\| 62.50<br>13/05 |
| 2008 | **9.4**<br>20.0\| 71.25<br>12/08 | **10.0**<br>14.0\| 92.50<br>02/05 | **7.0**<br>-9.0\| 62.50<br>22/07 | **11.5**<br>-19.0\| 51.25<br>17/02 | **13.9**<br>-35.0\|128.75<br>14/09 | **21.4**<br>-57.0\| 57.50<br>17/07 |
| 2009 | **10.0**<br>13.0\| 57.50<br>05/07 | **9.8**<br>17.0\| 87.50<br>20/07 | **7.4**<br>-9.0\| 48.75<br>02/07 | **10.1**<br>-22.0\| 53.75<br>08/02 | **14.1**<br>-35.0\| 27.50<br>24/06 | <u>**23.2**</u> (<u>ab.m</u>)<br><u>-49.0\| 86.25</u><br><u>21/09</u> |

Note. Results are presented in the following format: the first line is the extreme height $H_{S\max}$ in meters (**bold font**), the second line is the space coordinates (lattitude|longitude), the lowest line is the time(day/month). The sign minus means the southern hemisphere. Underlining means the greatest maximum of $H_{S\max}$ in the zone. Abbreviation "<u>ab.m</u>" means the absolute maximum of wave height in whole IO for all the period considered.

wind speed but by the fetch, as well. The said above increases the significance of the results for the problems of risk assessment for shipping and maritime activities [11]. Besides, the obtained values $H_{S\,max}$ are of considerable practical interest as one of the main characteristics of the wave-field regime in different zones of IO.

Additionally it is interesting to note that, in general, the location of extremes $H_{S\,max}$ in IO has much less chaotic distribution in space and time, compared with the field of $W_{max}$ [3]. A complete picture of the maxima of wave heights, built in the form of "synthetic" map $H_{S\,max}$ (x, y), the points of which are not tied to a single time, reflects the zone-structure of the spatial inhomogeneity of the mean waves field in the ocean (Fig. 2-app, see Appendix ).

*5.3. Histograms of the wave heights distribution (in zones and entire IO)*

The most complete probabilistic information about the wave height field is given by the histograms, the study of which has been performed in a lot of papers (see references in [9-11]). The histograms are the numerical representation of the probability density function (PDF). For a variety of geophysical fields, PDF is usually parameterized by a kind of Weibull distribution, the general form of which is given by formula[10]

$$P_{n,a}(W) = \frac{n}{a}\left(\frac{W}{a}\right)^{n-1} \exp\left[-\left(\frac{W}{a}\right)^n\right] \qquad (5.2)$$

where $W$ is the random variable, $n$ is the dimensionless exponent of the distribution, and $a$ is the "scale" parameter. In the case of stationary and pure (unmixed) wave field, the PDF for instantaneous wave heights is well described by the Rayleigh distribution [10], which represents the special case of (5.2) with exponent $n = 2$. However, it is important to note that we consider the PDF not for a random series of wave heights but for the statistical characteristics of the wave field: the significant wave height, $H_S$, given by formula (1.2). Therefore, we have no reason to expect a PDF close to the Raleigh kind. Just these differences in PDF constitute the main interest of our calculations.

It is known, however, that the parameters of distribution (5.2) depend strongly on the conditions of wave formation. Therefore, we should expect that on the scale of the ocean averaging and the 12-year period consideration, these parameters could vary greatly both in space and time, and depend on the scale of the space-time averaging of wave field. Such variability of probabilistic structure of the wave height field leads to a need of separate and direct calculation characteristics of parameterization (5.2) for each of areas and time periods under consideration. A full discussion of this problem requires a special discussion in further.

Here, we do not aim to test hypotheses about validity of the wave-height histograms parameterization in the form of (5.2). This is a separate and independent task. Instead, at this stage of research, we calculate the first 4 statistical moments for significant wave heights $H_S$ with using the distribution function $P(H_S)$ obtained from the histograms following from our calculations for wave fields.

Each of the statistical moments has a certain statistical sense. Thus, the first moment M = $<H_S> = \int H_S P(H_S) dH_S$, has a meaning of the average wave height. The second moment $<(H_S-M)^2>$ provides the integral information about a data scattering, the so-called dispersion of the distribution: $D = (<(H_S-M)^2>)^{1/2}$. In particular, the value of $D$ is important for determining a statistical significance of mean values estimates (the confidence interval for the mean value M is $\Delta_M = (D/N^{1/2}$, where $N$ is the length of the random series). The third moment, called the asymmetry and given as $A = <(H_S-M)^3>/D^3$, is a measure of deviation of empirical PDF from the PDF of the Gaussian type. Positive values of asymmetry $A$ means that the most probable value of $H_S$ is greater than the average one (and vice versa). And the fourth moment, called the excess and given as $E = \{<(H_S-M)^4>/D^4 - 3\}$, indicates the shape of peak and tail of the distribution. If $E > 0$, the peak of the distribution function is sharper than in the Gaussian distribution, what is accompanied by an increased probability of high values of wave heights, and vice versa.

Taking into account the main aim of our study, the histograms of wave heights are not of great interest for us. However, for the sake of completeness, here we give a series of graphical representations for certain histograms (Fig. 10), accompanying them by a brief analysis. In addition, we give quantities M, D, A, E, estimated at fixed points of zones, for the whole zones, and for the whole ocean (Tab. 4). Values of characteristics $n$ and $a$ for distributions (5.2), obtained from values of $M$ and $D$, are also included in Tab. 4, for completeness. The relevant formulas for estimating $n$ and $a$ are given in [3].

*5.3.1. Analysis of histograms.*

Histograms of significant wave height $H_S$, constructed with increment of 0.2m for whole 12-year period at the central points of the zones, are shown in Fig. 10a. It is seen a regular increase of the tail of PDF while shifting the area to the south. Irregularity of the graphs (especially in the northern zones) is due to the mixed nature of wave field (i.e. due to presence of swell). This irregularity is reduced significantly after spatial integrating of statistics over the whole zones (dashed lines for zones Z1 and Z6).

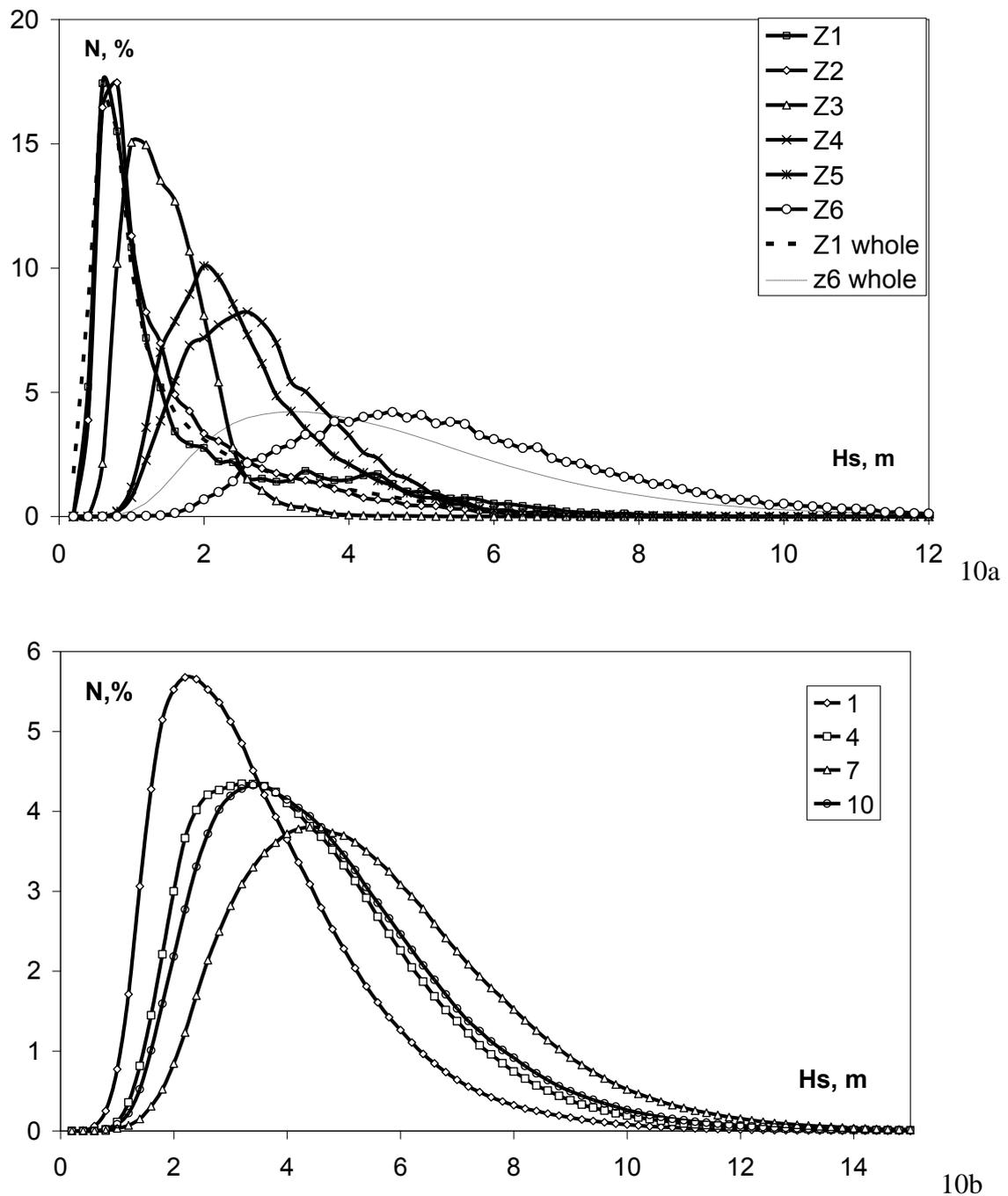

Fig. 10. Smoothed histograms for wave height $H_S$, obtained for the whole period:
a) for the central points of the zones; b) for the whole zone Z6 but for the certain four months (digits on right are the numbers of months).

In particular, in Fig. 10b the histogram are shown, integrated for the entire zone Z6 and obtained for different seasons (4 different months). Attention is called to intensification of the wave dynamics in the summer period (July). Besides, the spring and fall seasons show very close forms of PDF. In this respect, the PDF for the waves has the same dynamics as PDF for wind speed [3]. Apparently, this result is of the most interest in terms of histograms.

More detailed information follows from the estimates of spatial and temporal distribution of the first four statistical moments shown in Tab. 4.

Table 4.

Distribution of the statistical characteristics of the wave field $H_S(x,y,t)$ for the central points of zones, zones in whole(for seasons and all the period), and whole IO.

| Type of histogram | M, м | D, м | A | E | $n$ | $a$, м |
|---|---|---|---|---|---|---|
| $Z_1$ | 1.9 | 1.6 | 1.5 | 1.4 | 1.2 | 2.0 |
| $Z_2$ | 1.6 | 1.2 | 1.9 | 4.6 | 1.3 | 1.7 |
| $Z_3$ | 1.6 | 0.5 | 1.0 | 2.1 | 2.9 | 1.7 |
| $Z_4$ | 2.8 | 1.0 | 0.8 | 0.5 | 2.9 | 3.2 |
| $Z_5$ | 2.6 | 1.1 | 1.6 | 4.2 | 2.5 | 2.9 |
| $Z_6$ | 5.6 | 2.1 | 0.7 | 0.2 | 2.9 | 6.1 |
| Z1-w\|s | 1.0 \| 2.9 | 0.6 \| 1.5 | 1.5 \| 0.7 | 3.1 \| 0.3 | 1.8 \| 2.1 | 1.1 \| 3.3 |
| Z2-w\|s | 0.9 \| 2.3 | 0.5 \| 1.1 | 1.3 \| 1.2 | 2.1 \| 2.1 | 1.9 \| 2.3 | 1.0 \| 2.6 |
| Z3-w\|s | 1.2 \| 2.3 | 0.5 \| 0.7 | 2.0 \| 0.9 | 7.5 \| 1.9 | 2.4 \| 3.6 | 1.4 \| 2.5 |
| Z4-w\|s | 1.8 \| 3.2 | 0.8 \| 1.0 | 1.5 \| 0.7 | 3.9 \| 0.6 | 2.5 \| 3.5 | 2.0 \| 3.5 |
| Z5-w\|s | 2.1 \| 3.3 | 0.9 \| 1.4 | 1.6 \| 1.6 | 4.2 \| 3.7 | 2.6 \| 2.5 | 2.3 \| 3.7 |
| Z6-w\|s | 3.5 \| 5.5 | 1.7 \| 2.1 | 1.2 \| 0.6 | 1.7 \| 0.1 | 2.2 \| 2.8 | 4.0 \| 6.1 |
| Z1-tot | 1.6 | 1.3 | 1.7 | 2.7 | 1.2 | 1.7 |
| Z2-tot | 1.4 | 0.96 | 1.7 | 4.0 | 1.5 | 1.5 |
| Z3-tot | 1.6 | 0.74 | 1.1 | 1.9 | 2.3 | 1.8 |
| Z4-tot | 2.4 | 0.98 | 0.97 | 1.4 | 2.6 | 2.6 |
| Z5-tot | 2.7 | 1.2 | 1.7 | 4.7 | 2.3 | 3.0 |
| Z6-tot | 4.5 | 2.1 | 0.88 | 0.62 | 2.3 | 5.1 |
| IO-tot | 2.9 | 1.9 | 1.4 | 2.2 | 1.6 | 3.3 |

Note. The types of histograms: $Z_i$ is the histogram for the central point of the zone Zi; Zi-w|s is the histogram for whole zone Zi in the certain month for all the period(*w* means January, i.e. winter; *s* means July, i.e. summer); Zi-tot is the histogram for the whole zone and for all the period; IO-tot is the histogram for the whole IO for all the period.

*5.3.2. Analysis of the statistical moments for the wave field*

By analogy with the statistics for the wind field, the statistical moments distribution is presented in Tab. 4 for several different space-time scales of statistics integrating: statistics for values $H_S$ at the central points of zones; statistics calculated for each January and for each July with integrating for all years; integrated over the zones and for the entire period; and integrated over whole IO, as well.

For mean values of wave height $H_S$ it is seen their regular increase while shifting to the south what does not require any explanation. Herewith, the summer months in all zones of IO are characterized by a more rough sea.

Dispersion of wave heights, *D*, is rather small and does not show a clear latitudinal dependence. The smallness of the dispersion provides a small relative confidence interval for the averaged mean values (about 1%). This value characterizes the accuracy of the major part of estimates obtained in this paper.

Regarding to assessments of asymmetry *A* and kurtosis *E*, it should be noted the following. First, in summer season they are always less than in winter. Second, in winter they are always significant: asymmetry *A*, as a rule, is higher than 1, whilst excess *E* is often greater than 3, indicating a significant probability of occurrence of high waves. These quantities have the lowest values for summers in zone Z6, what is compensated by high values of *M* and scale-parameter of distribution *a* for this period.

With respect to distribution of index *n*, one should note that, despite a considerable scattering, the values of *n* are not very different from two for winter periods or for total time period, after averaging statistics over zones. However, these values, obtained by averaging over zones in the summer season, or at the central points of these zones for all 12 years, can exceed 3.5 units, indicating a significant deviation from the Rayleigh statistical distribution. A detailed discussion of this issue requires further investigations.

In this connection it is worthwhile to pay attention on the result of the moments assessment for whole IO (the last row in Tab. 4). From this it clearly follows the full unacceptability of study of the wave field statistics (apparently, the statistics of any geophysical fields) on the scale of whole ocean, because this approach does not reflect regional peculiarities of this field.

### 6. General inferences and conclusion

The most general inferences from the results of present work are as follows.

Firstly, it is shown that the field of waves, generated by the inhomogeneous wind field, is rather inhomogeneous in the area of the whole Indian Ocean. In spite of the nonlocal dependence of wave field on wind field, it was found that the partition of the wave field in area of IO can be linked to the previously adopted zoning for the wind field [3]. On this basis, the analysis of the variability scales for a wave field is conducted separately for each zone (see Section 4), whilst the values, averaged annually and across the ocean, are analyzed only in order to determine the 12-year wave-field variability (i.e. the long-term trends in wave heights and energy averaged over whole IO, section 5.1).

Second, the set of variability scales is found for the wave height values of different averaging: both for 3-hour series at the central points of zones and for series produced by their different spatial and temporal averaging. It was shown that they have significant difference for different zones. However, for each area, these sets of scales are robust to variations in types of averaging.

On the example of the spectral analysis for the wave series (Section 4), it was shows that the major scale of variability has a period of 1 year. On this background, there are variability with the period of 0.5year and a number of scales in the range of 40-10 days. Herewith, the period of 0.5 year is typical for northern zones Z1-Z3, only. For the southern zones Z4-Z6, it was established a weak variation of the wave-height spectra intensity in the range of periods from 100 to 10 days. This feature of the shape for the wave heights spectra repeats the similar feature of the wind speed spectra[3], which allows us to treat it (by analogy) as a manifestation of the lack of correlation for variability of wave fields in this range of scales.

In range of periods less than 10 days, for all scales of wave field averaging, there is a power-law decay of the wave heights spectrum of kind (4.1). More detailed conclusions about the variability scales and interpretation of the spectra shapes are presented in subsection 4.5.

Thirdly, we have shown (Sections 4 and 5.1) that the study of wave energy characteristics has an independent and justified interest. For the first time, we have found the wave-field energy store in IO, estimated for different scales of averaging. It was shown that the time of full pumping of waves by wind has in IO only a few days (actually, a 7-10days, section 4.4). Thus, the real time scale is established of an energy exchange between wind and wave fields, which is seen in the spectrum of the wave energy series.

Besides, in Section 5.1 the estimates of seasonal, interannual and 12-year variability are found for wave elevation and wave energy fields. For the first time it was found that the positive trend in wave heights and wave energy, averaged across the ocean, takes place for the time interval 1998-2009yy. The former is about 1% and the latter does of 2% per year. This result correlates well with the previously established trend of the average value of wind energy, having 1% per year [3]. This is the most impressive result of the paper.

In this regard, it should be noted that the most recent data, presented in article [9], show that the corresponding estimates of the trends for wind speed and wave heights, derived from the analysis of long-term satellite data averaged over whole World ocean, have values of 2-3 times below those obtained by us. This difference raises the question of clarification the extent of authenticity and degree of reliability both the remote sensing data and numerical simulations. Leaving this fundamentally important issue in prospective, note that it is this difference increases the significance of the estimates found here, the additional clarification of which becomes the necessary and urgent continuation of the work in this direction.

In conclusion we note that the processing the technique, proposed in [3] and used here for data of a large (geophysical) format, is proved to be highly informative. There is a hope that an expanding the range of periods and geography, used for analysis of the proposed set of geophysical fields (wind, waves and their energy), will helps to clarify and identify new patterns of their energy-exchange relationship at different scales, the nature of which is little understood yet.

The authors thank academician Golitsyn G.S. for numerical discussions of the points under investigation and the chief of the project Prof. Ginzburg A.S. for a permanent support in the work execution. We grateful to Mokhov I.I. for his constant interest to the topic. This work was supported by RFBR, grant № 10-05-92662-IND_a.

## Appendix

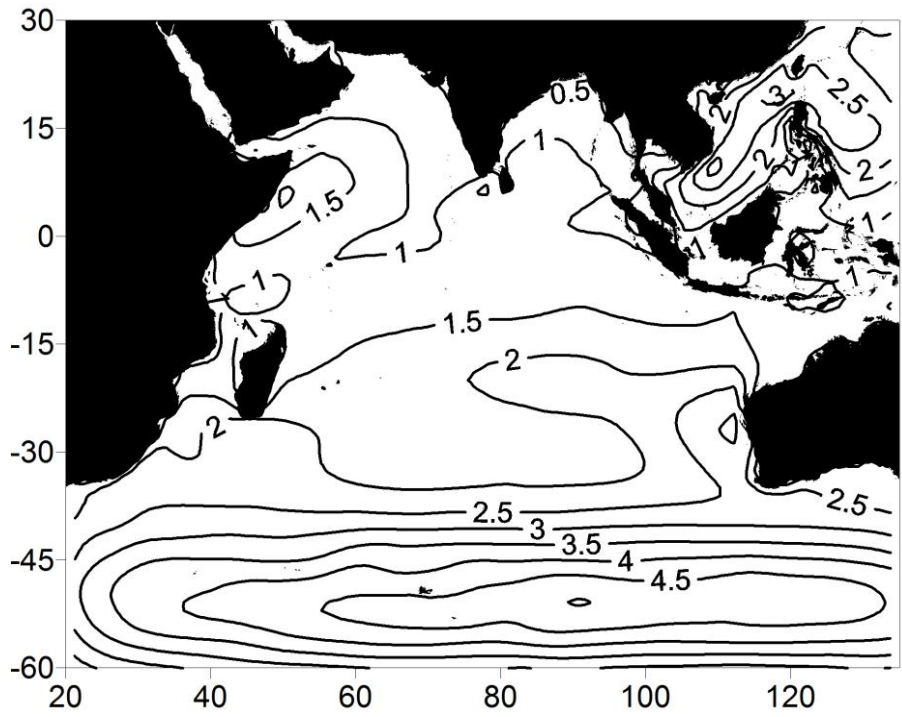

a)

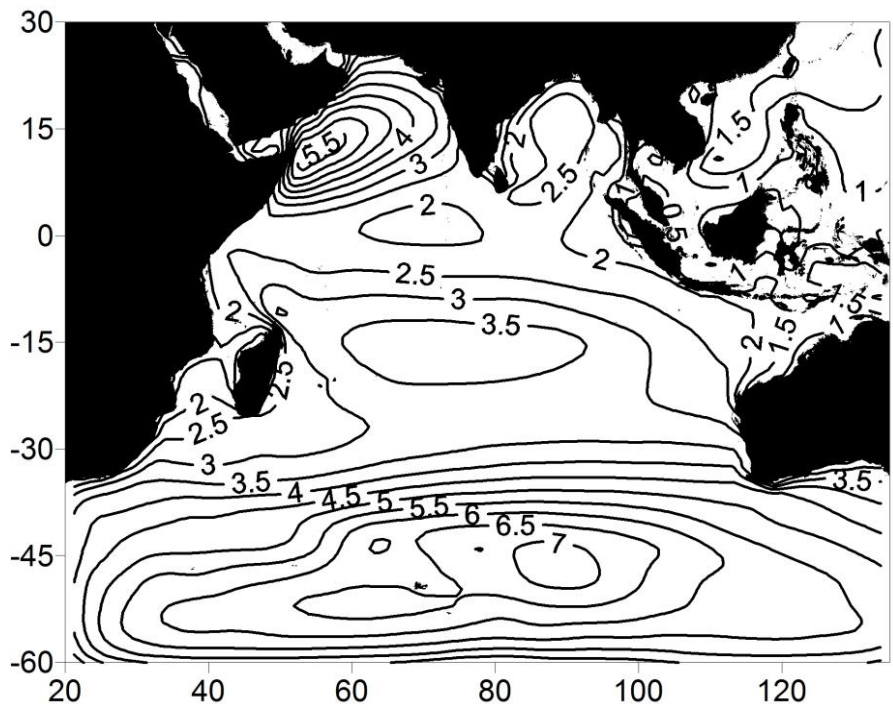

b)

Fig.-1pp. Seasonal maps of $<H_S(i,j,T)>$: a) mean January map, b) mean July map.

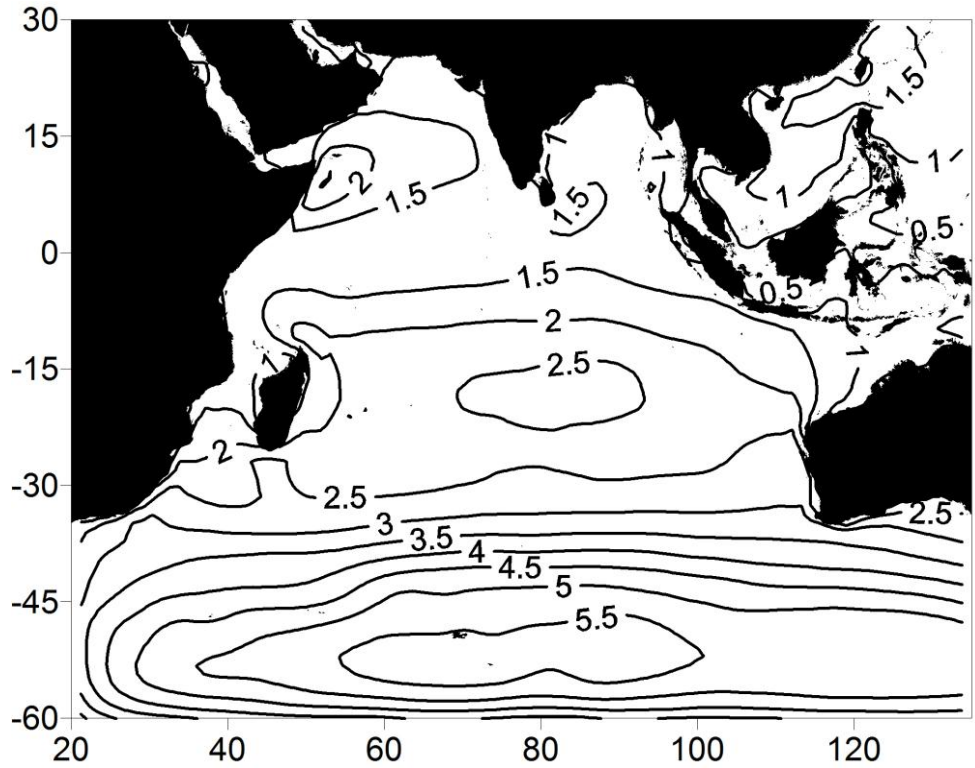

a)

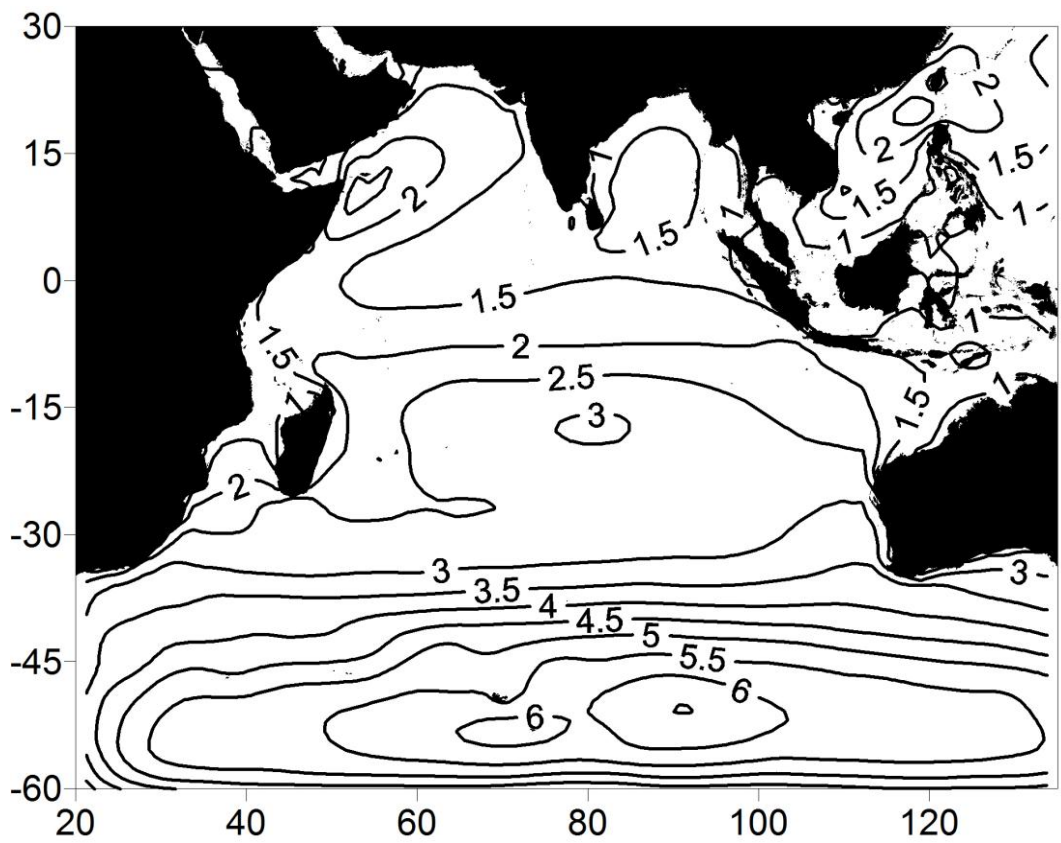

b)

Fig.-2pp. Annual maps of $<H_S(i,j,T)>$: a) mean 1998y map, b) mean 2008y map.

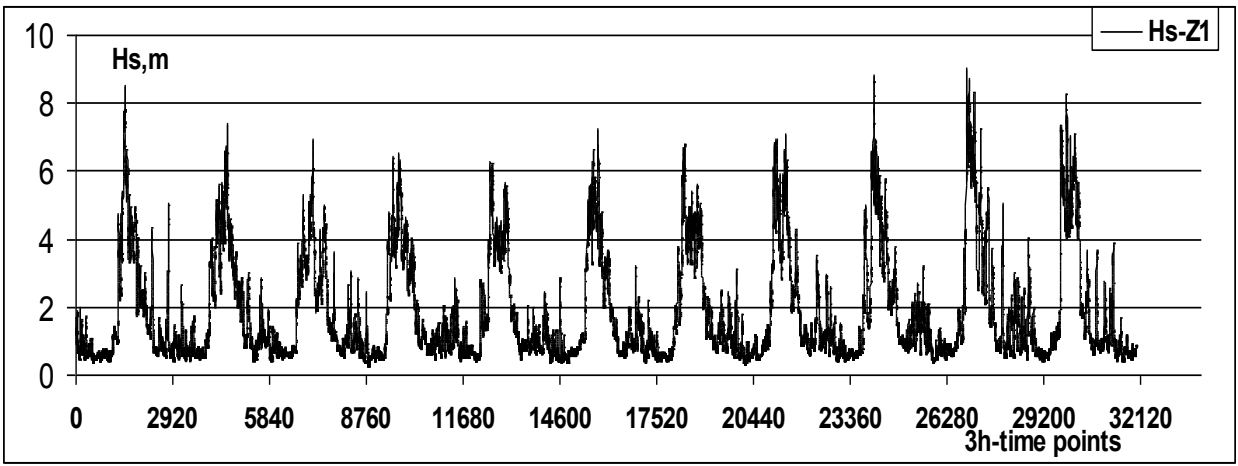

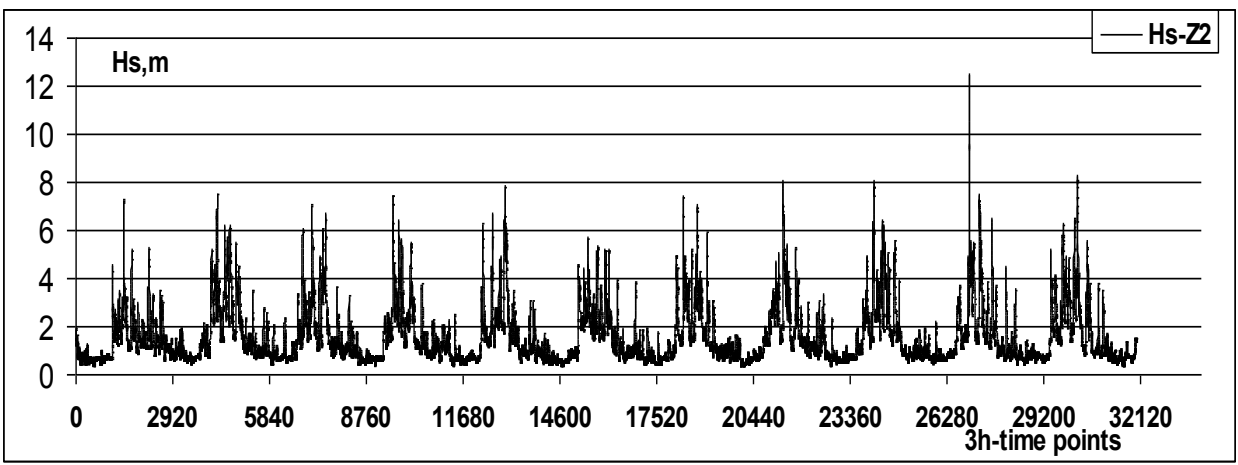

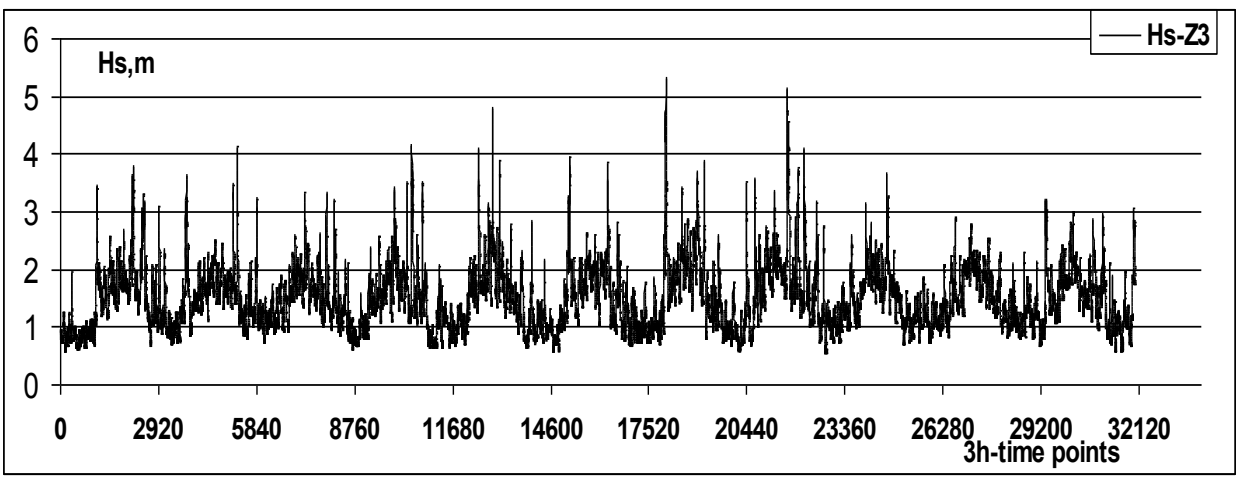

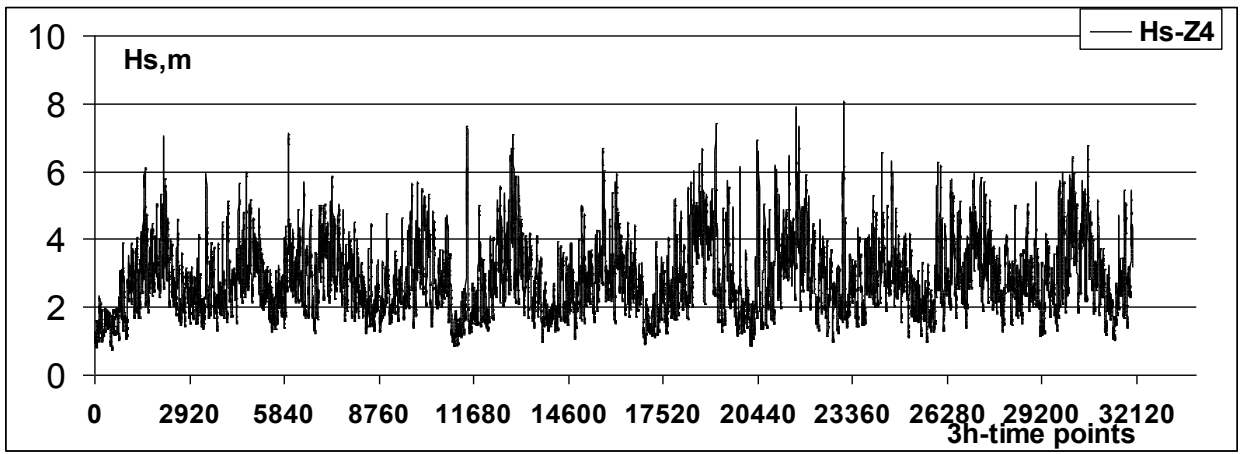

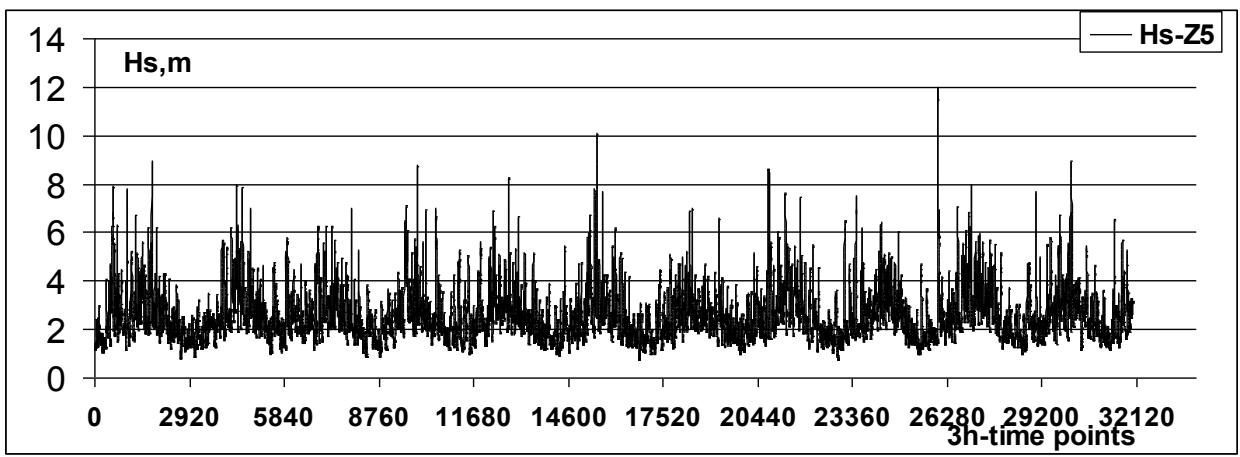

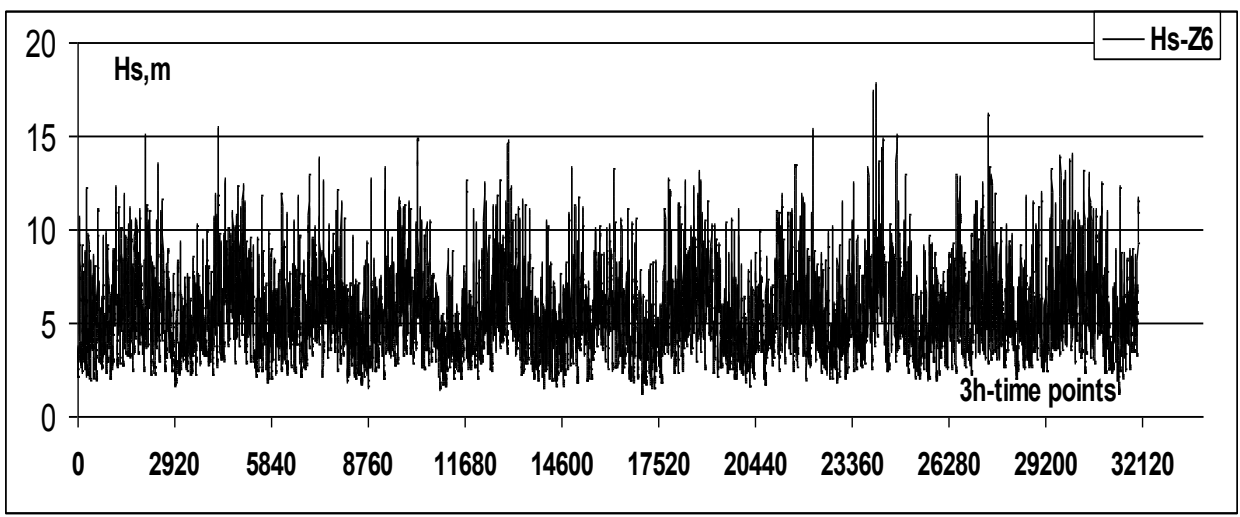

Fig. 3-app. The 3h-time history for wave heights in the central points of the zones in IO